\documentclass[a4paper,12pt]{article}
\usepackage{amsmath,amssymb,bm}
\usepackage{enumerate}
\usepackage{mathrsfs}
\usepackage{amsthm}

 \newtheorem{thm}{Theorem}
 \newtheorem{cor}{Corollary}
 \newtheorem{lem}{Lemma}
 \newtheorem{prop}{Proposition}
 \newtheorem{fact}{Fact}
 \newtheorem{step}{Step}
 \theoremstyle{definition}
 
 \theoremstyle{remark}

\numberwithin{equation}{section}

\title{On the Basis of the Hamilton-Jacobi-Bellman Equation in Economic Dynamics}
\author{Yuhki Hosoya\thanks{E-mail: hosoya(at)tamacc.chuo-u.ac.jp}\\Faculty of Economics, Chuo University\thanks{742-1 Higashinakano, Hachioji-shi, Tokyo 192-0393, Japan.}}
\date{\today}
\begin{document}
\maketitle

\begin{abstract}
We consider the classical Ramsey-Cass-Koopmans capital accumulation model and present three examples in which the Hamilton-Jacobi-Bellman (HJB) equation is neither necessary nor sufficient for a function to be the value function. Next, we present assumptions under which the HJB equation becomes a necessary and sufficient condition for a function to be the value function, and using this result, we propose a new method for solving the original problem using the solution to the HJB equation. Our assumptions are so mild that many macroeconomic growth models satisfy them. Therefore, our results ensure that the solution to the HJB equation is rigorously the value function in many macroeconomic models, and present a new solving method for these models.

\vspace{12pt}

\vspace{12pt}
\end{abstract}

\section{Introduction}
In modern macroeconomic dynamic models, the predominant approach is to describe economic behavior through variational problems. There is a reason for this. Lucas \cite{LU} submitted his famous analysis (Lucas critique) on the great divergence between macroeconomic theory and reality in the 1970s. This critique is to the effect that one should not use a theoretical model where changing economic policy based on this theory would make the original theory less applicable. In fact, by the mid-1970s, the negative correlation between inflation and unemployment, which was supposed to have been confirmed by Philips \cite{PH} using 100 years of data, had broken down, and this was regarded to be caused by the use of theories that could not withstand the Lucas critique.

The Lucas critique was regarded as a highly appropriate explanation for why reality is unpredictable by traditional models. Thus, in subsequent macroeconomic theory performing analysis using models that can withstand this critique has come to be important. The easiest way to avoid this critique is to endogenize people's behavior and build a model in which behavior automatically changes when economic policy changes. The variational problem is very convenient for this purpose.

Models for analyzing the economy through the variational problem existed before the Lucas critique. Ramsey \cite{RA} proposed such a model, which was popularized by Cass \cite{CA} and Koopmans \cite{KO}, who modified it to make it easier to analyze; the model is now called the Ramsey-Cass-Koopmans (RCK) model.\footnote{Because this model is famous, many macroeconomic textbooks introduce this model; see, for example, Acemoglu \cite{AC}, Barro and Sala-i-Martin \cite{BS1}, Blanchard and Fischer \cite{BF}, and Romer \cite{ROM}.} This model includes centralized and decentralized types. In the centralized model, the central government plans economic fluctuations, whereas in the decentralized model, the household and the firm independently optimize their objective functions, and prices are determined so that supply matches demand. Although these two models are quite different, it had been shown that in many cases, the same result is derived from them. Hence, in this paper, we consider only the centralized model.

In modern macroeconomics, stochastic fluctuations caused by shocks from outside the economy are added to such a variational problem. The real business cycle (RBC) model or the dynamic stochastic general equilibrium (DSGE) model are examples, which are frequently used in cutting-edge research of macroeconomic dynamics. However, it is usually too difficult to solve these models analytically, or even to find an approximate solution. Hence, a method of analysis that uses the Hamilton-Jacobi-Bellman (HJB) equation was developed. This equation has been used frequently in recent studies to analyze economic movements and obtain some information.\footnote{For example, see Achdou et al. \cite{ABLLM}}

However, there is a major problem with these studies. They almost unconditionally believe that the value function of a variational problem is the unique classical solution to the HJB equation. Mathematically, such a belief is too optimistic. The study of viscosity solutions in the HJB equation was initiated by Lions \cite{LI} and Crandall and Lions \cite{CL2}, but they considered problems with a finite time interval. Variational problems used in economics usually have an infinite time interval. To the best of our knowledge, such problems were first analyzed by Soner \cite{SO1, SO2}, but these problems are difficult to treat, and in some problems, the value function is not even a viscosity solution of the HJB equation, as discussed by Barles \cite{BA}, for example. Despite the existence of such a problem, many economists ignore it. In fact, there is a ``proof'' known to economists that links the HJB equation to the value function, and this ``proof'' leads them to believe that the value function is a solution to the HJB equation. For example, Malliaris and Brock \cite{MB} contains such a ``proof''. If this ``proof'' is correct, problems such as those above would only occur in fields far from economics, and not appear in economic models. However, this is not true.

In this paper, we first indicate that there are many models in which the above relationship between the HJB equation and the value function is broken, even within the scope of the RCK model, which is the most traditional variational problem in economics (Section 2). Among these problems are those for which there is no classical solution to the HJB equation in the first place; those for which the value function is a classical solution to the HJB equation but there are infinitely many other classical solutions; and those for which the value function is not a viscosity solution to the HJB equation but there are infinitely many classical solutions to the HJB equation. Thus, the above problem regarding the relationship between the solution to the HJB equation and the value function is serious for economists.

On the other hand, there are many books and papers that provide a foundation for the HJB equation in models where the time interval is infinite. Ch.3 of Bardi and Capuzzo-Dolcetta \cite{BCD} is an example of a book, and Da Lio \cite{DL}, Baumeister et al. \cite{BLS}, and Hermosilla and Zidani \cite{HZ} are examples of papers. However, to the best of our knowledge, none of these studies is applicable to typical applications used in economics. Specifically, most of them include conditions such as boundedness and Lipschitz condition in what is called the instantaneous utility function in economics. The most frequent example of the instantaneous utility function in an application in economics is a class of functions called the CRRA function, which includes the logarithmic function, and the assumptions in the above study would eliminate such a function.\footnote{See Section 4 for a rigorous definition of the CRRA function.}

Therefore, it is necessary to provide a foundation that allows economists to use the HJB equation safely in macroeconomic models. The purpose of this paper is to solve this problem. First, we derive the result that, under some economically valid assumptions, if the value function is finite, then it is a classical solution to the HJB equation (Proposition \ref{Prop3}). We then add a condition for the value function to be finite and derive conditions for the value function to be a classical solution to the HJB equation (Theorem \ref{Theorem1}). We go one step further and show that, under an additional condition, the value function is the unique classical solution to the HJB equation on a certain function space (Theorem \ref{Theorem2}). Taken together, these results constitute a mathematical basis for the use of the HJB equation to determine the value function in our economic model, under reasonable assumptions.

Our assumptions are sufficiently weak to allow us to treat a much wider variety of problems than the usual RCK model. First, we do not assume the Inada condition, nor do we assume its negation.\footnote{The Inada condition is defined in Section 2.} Thus, we can deal with models that satisfy the Inada condition, or conversely, we can deal with problems that do not satisfy the Inada condition, such as the AK model. We also do not assume that the speed of capital accumulation is linear with respect to consumption. Furthermore, we allow the instantaneous utility function to depend not only on consumption but also on capital accumulation. Thus, the results of this paper can be applied to a considerable number of macroeconomic dynamic models.

Additionally, in the process of deriving our results, we derive an approach to compute a solution to the original problem using the solution to the HJB equation. (Corollaries 1, 2) This result is useful even in the RCK model in the following sense. In the usual RCK model, the analysis uses phase diagrams for the simultaneous differential equation of the capital accumulation equations and the Euler equations. This system is semi-stable. By setting the initial value of consumption on a stable manifold, we can obtain a solution to this problem. However, because the system is semi-stable, if we apply usual methods for approximating the solution to differential equations such as the Runge-Kutta method, the approximation error will increase due to the unstable manifold, and the calculation will eventually fail. By contrast, under mild assumptions, we have derived a one-dimensional ordinary differential equation (ODE) for calculating the optimal capital accumulation, and in the case of the RCK model, its steady state is stable. Therefore, the above problem does not occur and a usual approximate method can be used to obtain an approximate solution.

We deal with a wide range of variational problems, including the RCK model and its variants, but not models involving stochastic shocks. Originally, we wanted to deal with stochastic variational problems involving DSGE models, but their analysis is so difficult that we could not do so in this paper. Some readers may consider our results inadequate in this respect. However, as the examples in Section 2 show, even within the RCK model, the mathematical basis for dealing with the HJB equation has so far been vague. Our results are not perfect, but we believe they are a worthwhile first step.

Although our model is a continuous-time model, our results are very similar to those of dynamic programming in discrete-time models. For example, Proposition \ref{Prop3} and Theorem \ref{Theorem1} correspond to the result that the value function solves the Bellman equation. Theorem \ref{Theorem2}, conversely, corresponds to the result that the solution to the Bellman equation is the value function. In the process of producing these results, we define a function $c^*(p,k)$, which corresponds to the policy function. Therefore, Corollaries 1 and 2 correspond to the result that one can derive the solution to the original problem using the policy function. These results are all in Ch.4 of Stokey and Lucas \cite{SL}, and our results can be seen as a continuous-time version of those.

Finally, we consider the logarithmic AK model as an example of applying our results. In this model, we can solve the HJB equation quite easily and use it to obtain a solution to the original problem. Although this solution has already been obtained by Barro and Sara-i-Martin \cite{BS1}, it is interesting in the sense that it is derived via a completely different path.

In Section 2, we present an example in which the HJB equation does not characterize the value function. In Section 3, we elaborate on the setup of our model and define several terms and assumptions. In Section 4, we discuss the main results. In Section 5, we explain a computational example and discuss several topics. In Section 6, we present concluding remarks.

\section{Motivation}
The purpose of this section is to demonstrate that there are many examples of variational problems in which either the value function does not satisfy the HJB equation or there are infinitely many classical solutions to the HJB equation, or both, although the problem is classified in the traditional RCK optimal growth model.

The RCK model is represented by the following problem:
\begin{align}
\max~~~~~&~\int_0^{\infty}e^{-\rho t}u(c(t))dt \nonumber\\
\mbox{subject to. }&~c(t)\mbox{ is locally integrable},\nonumber \\
&~k(t)\mbox{ is locally absolutely continuous},\nonumber \\
&~k(t)\ge 0,\ c(t)\ge 0\mbox{ for all }t\ge 0,\label{CE}\\
&~\dot{k}(t)=f(k(t))-c(t)\mbox{ a.e.},\nonumber \\
&~k(0)=\bar{k},\nonumber
\end{align}
where $\rho>0$ and $\bar{k}>0$. Throughout this section, we assume that $u(c)$ is bounded from below, and thus $\int_0^{\infty}e^{-\rho t}u(c(t))dt$ can be defined for any locally integrable function $c(t)$, although this value may become $+\infty$. Let $\bar{V}$ denote the value function of the above model. Note that, the HJB equation is the following differential equation.
\[\sup_{c\ge 0}\{(f(k)-c)V'(k)+u(c)\}-\rho V(k)=0.\]
We present the following facts.\footnote{We present the formal definitions of several technical terms, including the value function and classical and viscosity solutions to the HJB equation, in the next section. Because these facts are important in this paper, we present them before introducing the formal definitions.}

\begin{fact}\label{Fact1}
If $u(c)=c, f(k)=k$, and $\rho=2$, then the value function $\bar{V}$ is finite, concave, and not a viscosity solution to the HJB equation.
\end{fact}

\begin{proof}[{\bf Proof of Fact \ref{Fact1}}]
Let $V_1$ be the value function of (\ref{CE}) when $\rho=1$. We first show that $V_1(\bar{k})=\bar{k}$. Consider the following pair of functions:
\[k^*(t)\equiv \bar{k},\ c^*(t)\equiv \bar{k}.\]
Then, this pair $(k^*(t),c^*(t))$ satisfies all constraints of the problem (\ref{CE}). Suppose that $(k(t),c(t))$ also satisfies all constraints of the problem (\ref{CE}). Then,
\begin{align*}
\int_0^Te^{-t}(u(c^*(t))-u(c(t)))dt=&~\int_0^Te^{-t}(c^*(t)-c(t))dt\\
=&~\int_0^Te^{-t}((k^*(t)-k(t))-(\dot{k}^*(t)-\dot{k}(t)))dt\\
=&~\int_0^T\frac{d}{dt}[e^{-t}(k(t)-k^*(t))]dt\\
=&~e^{-T}(k(T)-k^*(T))\\
\ge&~-e^{-T}\bar{k}\to 0\mbox{ as }T\to \infty,
\end{align*}
which implies that $(k^*(t),c^*(t))$ is a solution to (\ref{CE}) if $\rho=1$, and thus,
\[V_1(\bar{k})=\int_0^{\infty}e^{-t}\bar{k}dt=\bar{k}.\]

Second, we show that $0\le \bar{V}(\bar{k})\le V_1(\bar{k})=\bar{k}$. Because $u(c)\ge 0$ for all $c\ge 0$, $\bar{V}(\bar{k})\ge 0$. Choose any pair $(k(t),c(t))$ that satisfies all constraints of problem (\ref{CE}). Then, because $u(c(t))\ge 0$ for all $t\ge 0$,
\[\int_0^{\infty}e^{-2t}u(c(t))dt\le \int_0^{\infty}e^{-t}u(c(t))dt\le V_1(\bar{k})=\bar{k},\]
and thus, $\bar{V}(\bar{k})\le \bar{k}$, as desired.

Therefore, $\bar{V}(k)$ is finite for all $k>0$. Third, we show that the function $\bar{V}$ is concave. Choose $k_1,k_2>0$ and $s\in [0,1]$, and let $k=(1-s)k_1+sk_2$. Suppose that a pair $(k_i(t),c_i(t))$ satisfies all constraints of the problem (\ref{CE}) with $\bar{k}=k_i$ for $i\in \{1,2\}$. Define $k(t)=(1-s)k_1(t)+sk_2(t),\ c(t)=(1-s)c_1(t)+sc_2(t)$. Then, $(k(t),c(t))$ satisfies all constraints of the problem (\ref{CE}) with $\bar{k}=k$. Therefore,
\[\bar{V}(k)\ge \int_0^{\infty}e^{-2t}u(c(t))dt=(1-s)\int_0^{\infty}e^{-2t}u(c_1(t))dt+s\int_0^{\infty}e^{-2t}u(c_2(t))dt.\]
Because $(k_i(t),c_i(t))$ is arbitrary,
\[\bar{V}(k)\ge (1-s)\bar{V}(k_1)+s\bar{V}(k_2),\]
which implies that $\bar{V}$ is concave.

Finally, we show that any concave function $V$ is not a viscosity solution to the HJB equation. Suppose that $V$ is a concave viscosity solution to the HJB equation. Then, $V$ is locally Lipschitz, and thus it is differentiable almost everywhere on the positive half-line, and the following equation is satisfied for almost all $k>0$:\footnote{See Subsection 3.4.}
\begin{equation}
\sup_{c\ge 0}\{(k-c)V'(k)+c\}-2V(k)=0.\label{HJBCE}
\end{equation}
Because the left-hand side becomes infinite if $V'(k)<1$, we have that $V'(k)\ge 1$ for almost all $k>0$. Then, the supremum of the left-hand side is attained at $c=0$. Therefore, (\ref{HJBCE}) can be modified to the following equation:
\[kV'(k)=2V(k).\]
By the Carath\'eodory-Picard-Lindel\"of uniqueness theorem for the solution,
\[V(k)=V(1)k^2.\]
Therefore, $V'(k)<1$ if $k>0$ is sufficiently small, which is a contradiction. Thus, such a $V$ is absent. In particular, $\bar{V}$ is not a viscosity solution to the HJB equation. This completes the proof.
\end{proof}

\begin{fact}\label{Fact2}
If $u(c)=c+\sqrt{c}$, $f(k)=k$, and $\rho=1$, then there exist infinitely many classical solutions to the HJB equation.
\end{fact}

\begin{proof}[{\bf Proof of Fact \ref{Fact2}}]
The HJB equation is
\[\sup_{c\ge 0}\{(k-c)V'(k)+c+\sqrt{c}\}-V(k)=0.\]
If $V'(k)\le 1$, then the left-hand side becomes infinite, and thus this equation is violated. Therefore, for any classical solution $V$ to the above equation, $V'(k)>1$ for all $k>0$, and the supremum of the left-hand side can be attained at $c=\frac{1}{4(V'(k)-1)^2}$. Thus,
\[kV'(k)+\frac{1}{4(V'(k)-1)}=V(k).\]
This is a variety of Clairaut's differential equation. The general solution to this equation is
\[V(k)=Ak+\frac{1}{4(A-1)},\]
where $A>1$ is some constant. The singular solution to this equation is\footnote{We can check that $\bar{V}(k)=k+\sqrt{k}$. Therefore, in this case, the value function is a solution to the HJB equation.}
\[V(k)=k+\sqrt{k}.\]
Therefore, the claim of this fact is verified. This completes the proof.
\end{proof}

\begin{fact}\label{Fact3}
If $u(c)=c$, $f(k)=\sqrt{k}$, and $\rho>0$, then there exist infinitely many classical solutions to the HJB equation, and the value function is finite but not a viscosity solution to the HJB equation.
\end{fact}

\begin{proof}[{\bf Proof of Fact \ref{Fact3}}]
First, define $g(k)=\rho k+\frac{1}{4\rho}$. Because $f(k)$ is concave and the graph for $g$ is a tangential line of that for $f$ at $k=\frac{1}{4\rho^2}$, $g(k)\ge f(k)$ for all $k\ge 0$. Consider the following problem:
\begin{align}
\max~~~~~&~\int_0^{\infty}e^{-\rho t}u(c(t))dt \nonumber\\
\mbox{subject to. }&~c(t)\mbox{ is locally integrable},\nonumber \\
&~k(t)\mbox{ is locally absolutely continuous},\nonumber \\
&~k(t)\ge 0,\ c(t)\ge 0\mbox{ for all }t\ge 0,\label{CE2}\\
&~\dot{k}(t)=g(k(t))-c(t)\mbox{ a.e.},\nonumber \\
&~k(0)=\bar{k},\nonumber
\end{align}
and let $\hat{V}$ be the value function of this problem. Define
\[k^*(t)=\bar{k},\ c^*(t)=\rho \bar{k}+\frac{1}{4\rho}.\]
Then, $(k^*(t),c^*(t))$ satisfies all constraints of (\ref{CE2}). If $(k(t),c(t))$ also satisfies all constraints of (\ref{CE2}), then
\begin{align*}
\int_0^Te^{-\rho t}(u(c^*(t))-u(c(t)))dt=&~\int_0^Te^{-\rho t}(c^*(t)-c(t))dt\\
=&~\int_0^Te^{-\rho t}[\rho (k^*(t)-k(t))-(\dot{k}^*(t)-\dot{k}(t))]dt\\
=&~\int_0^T\frac{d}{dt}[e^{-\rho t}(k(t)-k^*(t))]dt\\
=&~e^{-\rho T}(k(T)-k^*(T))\\
\ge&~-e^{-\rho T}\bar{k}\to 0\mbox{ as }T\to \infty,
\end{align*}
which implies that $(k^*(t),c^*(t))$ is a solution to (\ref{CE2}), and thus $\hat{V}(k)=k+\frac{1}{4\rho^2}$.

Second, suppose that $(k(t),c(t))$ satisfies all constraints of (\ref{CE}). Define
\[\tilde{c}(t)=g(k(t))-\dot{k}(t).\]
Then, $(k(t),\tilde{c}(t))$ satisfies all constraints of (\ref{CE2}), and $\tilde{c}(t)\ge c(t)$ for all $t\ge 0$. Therefore,
\[\hat{V}(k)\ge \int_0^{\infty}e^{-\rho t}u(\tilde{c}(t))dt\ge \int_0^{\infty}e^{-\rho t}u(c(t))dt,\]
which implies that $\bar{V}(k)\le \hat{V}(k)$ for all $k$. Because $u$ is nonnegative, $\bar{V}(k)\ge 0$, and thus $\bar{V}(k)$ is finite.

Third, choose $k_1,k_2>0$ and $s\in [0,1]$, and let $k=(1-s)k_1+sk_2$. Let $(k_i(t),c_i(t))$ satisfy all constraints of (\ref{CE}) with $\bar{k}=k_i$, where $i\in \{1,2\}$, and define
\[k(t)=(1-s)k_1(t)+sk_2(t),\ c(t)=f(k(t))-\dot{k}(t).\]
Because $f$ is concave,
\[c(t)\ge (1-s)c_1(t)+sc_2(t).\]
Therefore,
\[\bar{V}(k)\ge \int_0^{\infty}e^{-\rho t}u(c(t))dt\ge (1-s)\int_0^{\infty}e^{-\rho t}u(c_1(t))dt+s\int_0^{\infty}e^{-\rho t}u(c_2(t))dt,\]
which implies that $\bar{V}$ is concave. Because every concave function is locally Lipschitz, we conclude that $\bar{V}$ is absolutely continuous on every compact interval.

Fourth, suppose that $\bar{V}$ is a viscosity solution to the HJB equation. Because $\bar{V}$ is differentiable almost everywhere,
\[\sup_{c\ge 0}\{(\sqrt{k}-c)\bar{V}'(k)+c\}=\rho \bar{V}(k)\]
for almost all $k>0$. If $\bar{V}'(k)<1$, then the left-hand side becomes $+\infty$, which is a contradiction. Therefore, $\bar{V}'(k)\ge 1$ for almost all $k>0$, and thus
\[\sqrt{k}\bar{V}'(k)=\rho \bar{V}(k)\]
for almost all $k>0$. By the Carath\'eodory-Picard-Lindel\"of uniqueness theorem,
\[\bar{V}(k)=Ae^{2\rho\sqrt{k}-2\rho},\]
where $A=\bar{V}(1)\ge 1$. However, because $\bar{V}'(k)=\frac{\rho \bar{V}(k)}{\sqrt{k}}$,
\[\lim_{k\to 0}\bar{V}'(k)=\lim_{k\to \infty}\bar{V}'(k)=+\infty,\]
which contradicts the concavity of $\bar{V}$. Therefore, $\bar{V}$ is not a viscosity solution to the HJB equation.

Finally, consider the function
\[V(k)=Ae^{2\rho\sqrt{k}-2\rho},\]
where $A>0$. Then,
\[V'(k)=\frac{\rho V(k)}{\sqrt{k}},\]
and thus
\[\lim_{k\to 0}V'(k)=\lim_{k\to \infty}V'(k)=+\infty.\]
This implies that the minimum of the $V'(k)$ is positive, and linear in $A$. Therefore, if $A$ is sufficiently large, then $V'(k)\ge 1$ for all $k>0$, and thus $V$ is a classical solution to the HJB equation. This completes the proof.
\end{proof}

\vspace{12pt}
\noindent
\textbf{What is wrong in the above counterexamples?}

In related studies, the reason why the value function $\bar{V}(\bar{k})$ solves the HJB equation is told as follows. First, for every $t>0$, we can obtain the following equation:
\[V(\bar{k})=\sup_{k(s),c(s)}\left\{\int_0^te^{-\rho s}u(c(s))ds+e^{-\rho t}\bar{V}(k(t))\right\}.\]
We can transform this equation into
\[\sup_{k(s),c(s)}\left\{\int_0^te^{-\rho s}u(c(s))ds+e^{-\rho t}\bar{V}(k(t))-e^{-\rho 0}\bar{V}(k(0))\right\}=0.\]
If we assume that $\bar{V}$ is differentiable and $c(s)$ is continuous, then the composition $\bar{V}\circ k$ is differentiable at $s=0$, and thus
\begin{align}
0=&~\sup_{k(s),c(s)}\left\{\int_0^te^{-\rho s}u(c(s))ds+e^{-\rho t}\bar{V}(k(t))-e^{-\rho 0}\bar{V}(k(0))\right\}\nonumber\\
=&~\sup_{k(s),c(s)}\left\{t\times \frac{d}{dT}\left.\left[\int_0^Te^{-\rho s}u(c(s))ds+e^{-\rho T}\bar{V}(k(T))\right]\right|_{T=0}+o(t)\right\}\label{DOUBT}\\
=&~\sup_{k(s),c(s)}\{t[u(c(0))-\rho \bar{V}(k(0))+\bar{V}'(k(0))\dot{k}(0)]+o(t)\}.\nonumber
\end{align}
Because $k(0)=\bar{k}$ and $\dot{k}(0)=f(\bar{k})-c(0)$, the last line depends only on $c(0)$, and thus,
\[0=\sup_c\{t[(f(\bar{k})-c)\bar{V}'(\bar{k})+u(c)]+o(t)\}-\rho \bar{V}(\bar{k})t.\]
Dividing the right-hand side by $t$ and letting $t\to 0$, we obtain the HJB equation.

However, we have already obtained counterexamples in Facts \ref{Fact1} and \ref{Fact3}. Therefore, the above ``proof'' is somewhat incorrect. Upon closer inspection, we find that there are at least two gaps in the above ``proof''.

First, we said that ``the last line of (\ref{DOUBT}) depends only on $c(0)$''. However, in the ``last line'', there is Landau's little-o symbol $o(t)$. To be precise, this line just summarizes all the parts that depend on something other than $c(0)$ in the form of $o(t)$, and $o(t)$ still depends on the form of the functions $k(s)$ and $c(s)$. Additionally, the `$o(t)$' part is $o(t)$ if we fix a pair of functions $k(s)$ and $c(s)$, but when we take the supremum over such functions, we are not really sure if it is actually $o(t)$.

Second, for the derivation of (\ref{DOUBT}), we made some assumptions that guarantee the differentiability of $V\circ k$ at $s=0$. Specifically, we assumed the differentiability of $V$ and the continuity of $c(s)$. However, in our model, $c(s)$ may be discontinuous, and there is no guarantee that $V$ is differentiable. Therefore, it is uncertain whether we can derive the equation (\ref{DOUBT}) in the first place.

We cannot determine which of these two gaps is more critical. However, we know that both gaps are almost never a problem when there is a solution to the original problem. The first gap seems to be a problem that can be solved using a solution $(k^*(s),c^*(s))$ to evaluate $o(t)$. Moreover, if a solution exists, then it will satisfy the Euler equation, and thus the optimal consumption path $c^*(s)$ will be continuous in many cases. Furthermore, the ``inner solution'' condition of Benveniste and Scheinkman \cite{BS2} will be satisfied in most models, and therefore their Theorem \ref{Theorem2} can be applied and the value function will be differentiable. Hence, we think that if the original problem has a solution, both gaps will be filled in most cases. In other words, we can imagine that the problem in Facts \ref{Fact1} and \ref{Fact3} probably arises because there is no solution to the original problem.

So, does this solve our problem? We think not. This is because, in many studies that apply the HJB equation, the solution to the original problem is not guaranteed to exist at the stage of using the HJB equation. Moreover, another problem that there may be more than one solution to the HJB equation (Facts \ref{Fact2} and \ref{Fact3}) has not been solved. In many cases, we want to obtain a value function by solving the HJB equation, and what is needed for this is not that the value function solves the HJB equation, but rather that a solution to the HJB equation is actually the value function. In the case of Fact \ref{Fact3}, it is clear that we cannot obtain the value function by solving the HJB equation because the value function is not a solution to the HJB equation. In contrast, Fact \ref{Fact2} tells us that we cannot obtain the value function by solving the HJB equation even if we can show that the value function is a solution to the HJB equation.

Thus, many studies that use the HJB equation are in a very precarious position. A way to solve this problem is first to consider the solution to the HJB equation as a ``candidate'' of the value function, rather than immediately considering it as the value function, and then to prove that this ``candidate'' is indeed the value function. If we follow this procedure, we can use the HJB equation appropriately. However, it is uncertain whether we can follow the above procedure at any time. Perhaps proving this ``candidate'' is really the value function may be tremendously difficult.

Another way to solve this problem is to prove in a general framework that a solution to the HJB equation is the value function under several assumptions. In this paper, we address this issue, and show that the value function is a solution to the HJB equation, and that there is no other solution provided we do not deviate from the general assumptions used in macroeconomic theory. This result will give many studies a credible foundation and greatly increase their persuasiveness.

\section{Model and Definitions}
\subsection{Model}
In this paper, we use the following notation. First, we define $\mathbb{R}^n_+=\{x\in \mathbb{R}^n|x_i\ge 0\mbox{ for all }i\}$ and $\mathbb{R}^n_{++}=\{x\in \mathbb{R}^n|x_i>0\mbox{ for all }i\}$. If $n=1$, we omit this notation and simply write these sets as $\mathbb{R}_+$ and $\mathbb{R}_{++}$, respectively.

The model that we discuss in this paper is the following.\footnote{The statement ``$\int_0^{\infty}e^{-\rho t}u(c(t),k(t))dt$ can be defined'' admits that the value of this integral may be $\pm \infty$.}
\begin{align}
\max~~~~~&~\int_0^{\infty}e^{-\rho t}u(c(t),k(t))dt \nonumber \\
\mbox{subject to. }&~c(\cdot)\in W,\nonumber \\
&~k(t)\mbox{ is locally absolutely continuous},\nonumber \\
&~k(t)\ge 0,\ c(t)\ge 0,\label{MODEL}\\
&~\int_0^{\infty}e^{-\rho t}u(c(t),k(t))dt\mbox{ can be defined},\nonumber \\
&~\dot{k}(t)=F(k(t),c(t))\mbox{ a.e.},\nonumber \\
&~k(0)=\bar{k},\nonumber
\end{align}
where $W$ denotes some set of functions. Note that, if the instantaneous utility function $u(c,k)$ is independent of $k$ and the technology function $F(k,c)=f(k)-c$, then the problem (\ref{MODEL}) coincides with the traditional RCK model (\ref{CE}). In the context of the RCK model, the requirement $\mathbb{R}_{++}\subset f'(\mathbb{R}_{++})$ (called the {\bf Inada condition}) is sometimes used. However, in this paper, we do not need this condition.

We make several assumptions for the above model. First, let $W_1$ be the set of all functions $c:\mathbb{R}_+\to\mathbb{R}_+$ such that it is locally integrable, and $W_2$ be the set of all functions $c:\mathbb{R}_+\to\mathbb{R}_+$ such that it is measurable and locally bounded. Note that $W_2\subset W_1$.

\vspace{12pt}
\noindent
\textbf{Assumption 1}. $\rho>0$, and $W$ is either $W_1$ or $W_2$.

\vspace{12pt}
\noindent
\textbf{Assumption 2}. The instantaneous utility function $u:\mathbb{R}^2_+\to \mathbb{R}\cup \{-\infty\}$ is a continuous and concave function on $\mathbb{R}^2_+$. Moreover, $u(c,k)$ is nondecreasing on $\mathbb{R}^2_+$, and increasing in $c$ and continuously differentiable on $\mathbb{R}^2_{++}$.\footnote{Because $u$ is increasing in $c$ on $\mathbb{R}^2_{++}$, if $(c,k)\in \mathbb{R}^2_{++}$, then $u(c,k)>u(2^{-1}c,k)\ge -\infty$. Therefore, $u(c,k)\in \mathbb{R}$ for every $(c,k)\in \mathbb{R}^2_{++}$.} Furthermore, there exists $c>0$ such that $u(c,0)>-\infty$.

\vspace{12pt}
\noindent
\textbf{Assumption 3}. The technology function $F:\mathbb{R}^2_+\to \mathbb{R}$ is a continuous and concave function that satisfies $F(0,0)=0$. Moreover, $F$ is decreasing in $c$, and there exist $d_1>0$ and an increasing function $\delta_2:\mathbb{R}_+\to\mathbb{R}_+$ such that $\delta_2(0)=0$ and $F(k,c)>-d_1k-\delta_2(c)$ for every $(k,c)$ such that $k>0$ and $c\ge 0$. If $W=W_1$, then there exists $d_2\ge 0$ such that $\delta_2(c)=d_2c$ for all $c\ge 0$. Furthermore, for every $c\ge 0$, there exists $k>0$ such that $F(k,c)>F(0,c)$.\footnote{The last assumption means that for every $\bar{c}>0$, there exists $k^*>0$ such that for every $c\in [0,\bar{c}]$, $k\mapsto F(k,c)$ is increasing on $[0,k^*]$. In fact, we can choose
\[k^*=\min_{c\in [0,\bar{c}]}\min\arg\max\{F(k,c)|0\le k\le 1\}.\]}

\vspace{12pt}
\noindent
\textbf{Assumption 4}. The function $\frac{\partial u}{\partial c}(c,k)$ is decreasing in $c$ on $\mathbb{R}^2_{++}$. Moreover, $\lim_{c\to 0}\frac{\partial u}{\partial c}(c,k)=+\infty$ and $\lim_{c\to \infty}\frac{\partial u}{\partial c}(c,k)=0$ for every $k>0$. Furthermore, for every $k>0$ and $M>0$, the function $c\mapsto \frac{\partial u}{\partial k}(c,k)$ is bounded on $]0,M]$.

\vspace{12pt}
\noindent
\textbf{Assumption 5}. The technology function $F$ is continuously differentiable in $c$ on $\mathbb{R}^2_{++}$.

\subsection{Interpretation of Our Model}
There are two types of optimal capital accumulation models: centralized and decentralized. In this paper, we deal with a centralized model, which is connected to the corresponding decentralized model by the fundamental theorem of welfare economics. In the centralized macroeconomic dynamic model, $k(t)$ represents the capital stock and $c(t)$ represents the consumption, and both are assumed to be non-negative. Some models assume that $k(t)$ is continuously differentiable and $c(t)$ is continuous, whereas others admit ``jump'' on the trajectory of $c(t)$. In this paper, we adopt the latter position, and thus only assume that $c(t)$ is either integrable or measurable and bounded on any compact interval. As a result, we cannot assume continuous differentiability for the capital stock either, and only assume that $k(t)$ is absolutely continuous on any compact interval.

As previously mentioned, the model (\ref{MODEL}) includes the RCK model, which is the most common in the literature on optimal capital accumulation models. In the RCK model, an instantaneous utility function $u(c)$ is given, which does not depend on $k$. Because $u(c)$ may be a logarithmic function $\log c$, boundedness of $u$ is not usually assumed, and $u(0)=-\infty$ is allowed. On the other hand, $u'>0, u''<0$, $\lim_{c\to 0}u'(c)=+\infty$, and $\lim_{c\to \infty}u'(c)=0$ are usually assumed. With a little checking, readers will notice that this model satisfies our Assumptions 2 and 4.

Next, in the RCK model, the relationship between the capital stock and consumption is determined as follows:
\[\dot{k}(t)=f(k(t))-c(t).\]
There are two economic relationships behind this equation. First, the function $g$ is called the production function, and it indicates that if the capital stock is $k$, then the output is $g(k)$. This output is separated into consumption and investment. If we let $c$ denote the consumption and $i$ denote the investment, we obtain the first relationship:
\[g(k(t))=c(t)+i(t).\]
On the other hand, the speed of increase of the capital stock $\dot{k}(t)$ is equal to $i(t)$ minus capital depreciation $dk(t)$, where $d\ge 0$. Thus, we obtain the second relationship:
\[\dot{k}(t)=i(t)-dk(t).\]
By connecting these two relationships, we have that
\[\dot{k}(t)=g(k(t))-dk(t)-c(t),\]
and if we define $f(k)=g(k)-dk$, then we obtain the desired relationship.

Usually, it is assumed that $g$ is concave, increasing, and continuous, $g(0)=0$, and there exists $k>0$ such that $g(k)>dk$. By setting $F(k,c)=g(k)-dk-c$, $d_1=d$, and $\delta_2(c)=c$, readers can check that this model satisfies our Assumptions 3 and 5. Thus, our model (\ref{MODEL}) includes the RCK model. Moreover, unlike many other studies, our assumptions do not assume either differentiability of the production function $f$ or the Inada condition. Therefore, the scope of our model is sufficiently wide at this point.\footnote{Because $g'(k)<d$ can occur in the RCK model, we do not assume that $F(k,c)$ is increasing in $k$ in Assumption 3.}

Clearly, we can also deal with many models that are not RCK models. First, we can deal with a model in which the capital stock increases the instantaneous utility.\footnote{Such models appear in, for example, Zou \cite{ZO} and Futagami and Shibata \cite{FS}.} One typical example of such a utility is $u(c,k)=(c^{\sigma}+k^{\sigma})^{\frac{A}{\sigma}}$, where $0<A<1$ and $\sigma<1, \sigma\neq 0$. This is a variant of the so-called CES utility function.\footnote{If $\sigma<0$, then we define $u(c,0)=u(0,k)=0$. Note that, if $A=1$, then $\lim_{c\to \infty}\frac{\partial u}{\partial c}(c,k)=1$ when $\sigma>0$, and $\lim_{c\to 0}\frac{\partial u}{\partial c}(c,k)=1$ when $\sigma<0$, and thus Assumption 4 is violated in any case.}

We should explain the assumption of the existence of $c>0$ such that $u(c,0)>-\infty$. Economically, this assumption means that a lack of capital stock is not fatal to people, and only a lack of consumption can be fatal. Mathematically, this assumption is needed for Step 5 of the proof of Proposition \ref{Prop3}. Note that, this requirement prohibits some functions, such as $u(c,k)=\log c+\log k$.

Second, we can deal with different technology structures from that in the RCK model. For example, we can handle a model in which too high consumption has a strong negative impact on the growth rate of $k$, and a model in which choosing too high investment cannot be digested at once and used for production. The former model is given by the function $F(k,c)=f(k)-h(c)$, where $h$ is an increasing, continuously differentiable, and convex function that satisfies $h(0)=0$ and has a Lipschitz constant $d_2$. The latter model is given by the function $F(k,c)=h(g(k)-c)-dk$, where $h$ is an increasing, continuously differentiable, and concave function that satisfies $h(0)=0$ and has a Lipschitz constant $d_2$. In both cases, it is easy to prove that our Assumptions 3 and 5 are satisfied. If we choose $W_2$ as $W$, then the existence of the Lipschitz constants can be removed, and thus we can treat many technologies such as $F(k,c)=f(k)-c^2$.\footnote{This function $F(k,c)=f(k)-c^2$ is obviously not able to treat if $W=W_1$, since $(c(t))^2$ may be not locally integrable even if $c(t)$ is locally integrable.}

Finally, we note that in Facts 1-3, (\ref{CE}) satisfies Assumptions 1-3 and 5. Only Assumption 4 is violated.

\subsection{Admissibility and the Value Function}

We say that a pair of real-valued functions $(k(t),c(t))$ defined on $\mathbb{R}_+$ \textbf{admissible} if $k(t)$ is absolutely continuous on every compact interval, $c(t)\in W$, $k(t)\ge 0, c(t)\ge 0$, $\int_0^{\infty}e^{-\rho t}u(c(t),k(t))dt$ can be defined, and
\begin{equation}
\dot{k}(t)=F(k(t),c(t))\ \mbox{a.e.}.\label{TC}
\end{equation}
Note that, if $k(t)$ is absolutely continuous on every compact interval, it is differentiable almost everywhere and $\int_a^b\dot{k}(t)dt=k(b)-k(a)$ for all $a,b$ with $0\le a<b$.

Let $A_{\bar{k}}$ denote the set of all admissible pairs such that $k(0)=\bar{k}$. Using the notation of $A_{\bar{k}}$, we can simplify the model (\ref{MODEL}) as follows:
\begin{align*}
\max~~~~~&~\int_0^{\infty}e^{-\rho t}u(c(t),k(t))dt\\
\mbox{subject to. }&~(k(t),c(t))\in A_{\bar{k}}.
\end{align*}
For each $\bar{k}>0$, let
\[\bar{V}(\bar{k})=\sup\left\{\left.\int_0^{\infty}e^{-\rho t}u(c(t),k(t))dt\right| (k(t),c(t))\in A_{\bar{k}}\right\}.\]
We call this function $\bar{V}$ the \textbf{value function} of the problem (\ref{MODEL}).\footnote{We can easily show that $A_{\bar{k}}$ is nonempty for all $\bar{k}\ge 0$. Therefore, we can define $\bar{V}$ on $\mathbb{R}_+$. However, in this paper, we consider that the domain of $\bar{V}$ is $\mathbb{R}_{++}$ for some technical reasons, and thus $\bar{V}(0)$ is not treated throughout this paper except Subsection 5.2, although it can be defined.}

We call a pair $(k^*(t),c^*(t))\in A_{\bar{k}}$ a \textbf{solution} if and only if the following two requirements hold. First,
\[\int_0^{\infty}e^{-\rho t}u(c^*(t),k^*(t))dt\in \mathbb{R}.\]
Second, for every pair $(k(t),c(t))\in A_{\bar{k}}$,
\[\int_0^{\infty}e^{-\rho t}u(c^*(t),k^*(t))dt\ge \int_0^{\infty}e^{-\rho t}u(c(t),k(t))dt.\]
These two requirements can be summarized by the following formula:
\[\int_0^{\infty}e^{-\rho t}u(c^*(t),k^*(t))dt=\bar{V}(\bar{k})\in \mathbb{R}.\]
Note that, for all $(k(t),c(t))\in A_{\bar{k}}$ such that $\int_0^{\infty}e^{-\rho t}u(c(t),k(t))dt\in \mathbb{R}$ and $T>0$, if $k(T)>0$, then
\begin{align*}
\int_0^Te^{-\rho t}u(c(t),k(t))dt=&~\int_0^{\infty}e^{-\rho t}u(c(t),k(t))dt-\int_T^{\infty}e^{-\rho t}u(c(t),k(t))dt\\
\ge&~\int_0^{\infty}e^{-\rho t}u(c(t),k(t))dt-e^{-\rho T}\bar{V}(k(T)).
\end{align*}
In particular, if $\int_0^{\infty}e^{-\rho t}u(c(t),k(t))dt>M$, then for all $T>0$ such that $k(T)>0$,
\[\int_0^Te^{-\rho t}u(c(t),k(t))dt>M-e^{-\rho T}\bar{V}(k(T)).\]
Therefore, if $\bar{V}(\bar{k})$ is finite, then for every $\varepsilon>0$, there exists a pair $(k(t),c(t))\in A_{\bar{k}}$ such that either $k(T)=0$ or
\[\int_0^Te^{-\rho t}u(c(t),k(t))dt>\bar{V}(\bar{k})-e^{-\rho T}\bar{V}(k(T))-\varepsilon\]
for every $T>0$.

\subsection{The HJB Equation}
The HJB equation is given as follows.
\begin{equation}
\sup_{c\ge 0}\{F(k,c)V'(k)+u(c,k)\}-\rho V(k)=0.\label{HJB}
\end{equation}
A function $V:\mathbb{R}_{++}\to \mathbb{R}\cup \{-\infty\}$ is called a \textbf{classical solution} to the HJB equation if and only if $V$ is continuously differentiable and equation (\ref{HJB}) holds for every $k>0$.

It is known that, in many models of the dynamic control problem, there exists no classical solution to the HJB equation. Hence, we should extend the notion of the solution. First, a function $V:\mathbb{R}_{++}\to \mathbb{R}$ is called a \textbf{viscosity subsolution} to (\ref{HJB}) if and only if it is upper semi-continuous, and for every $k>0$ and every continuously differentiable function $\varphi$ defined on a neighborhood of $k$ such that $\varphi(k)=V(k)$ and $\varphi(k')\ge V(k')$ whenever $k'$ is in the domain of $\varphi$,
\[\sup_{c\ge 0}\{F(k,c)\varphi'(k)+u(c,k)\}-\rho V(k)\le 0.\]
Second, a function $V:\mathbb{R}_{++}\to \mathbb{R}$ is called a \textbf{viscosity supersolution} to (\ref{HJB}) if and only if it is lower semi-continuous, and for every $k>0$ and every continuously differentiable function $\varphi$ defined on a neighborhood of $k$ such that $\varphi(k)=V(k)$ and $\varphi(k')\le V(k')$ whenever $k'$ is in the domain of $\varphi$,
\[\sup_{c\ge 0}\{F(k,c)\varphi'(k)+u(c,k)\}-\rho V(k)\ge 0.\]
If a continuous function $V:\mathbb{R}_{++}\to \mathbb{R}$ is both a viscosity sub- and supersolution to (\ref{HJB}), then $V$ is called a \textbf{viscosity solution} to (\ref{HJB}).

Suppose that $V$ is a viscosity solution to the HJB equation and is differentiable at $k>0$. Then, it is known that
\[\sup_{c\ge 0}\{F(k,c)V'(k)+u(c,k)\}-\rho V(k)=0.\]
See Proposition 1.9 of Ch.2 of Bardi and Capuzzo-Dolcetta \cite{BCD}.

\subsection{Subdifferentials and Left- and Right-Derivatives}
In this paper, we heavily use subdifferential calculus. We introduce the notion of the subdifferential and several results. For the proofs of these results, see textbooks on convex analysis, such as Rockafeller \cite{ROC}.

Suppose that a function $G:U\to \mathbb{R}$ is concave, $U\subset \mathbb{R}^n$ is convex, and the interior $V$ of $U$ is nonempty. Choose any $x\in V$. We define
\[\partial G(x)=\{p\in \mathbb{R}^n|G(y)-G(x)\le p\cdot (y-x)\mbox{ for all }y\in U\}.\]
Then, we can show that $\partial G(x)$ is nonempty. The set-valued mapping $\partial G$ is called the \textbf{subdifferential} of $G$.\footnote{Formally, the subdifferential is defined for not concave but \textbf{convex} functions, and thus the inequality in the definition is reversed. In this view, the name `subdifferential' may be not appropriate, and `superdifferential' may be more suitable. However, in the literature of economics, these two notions are not distinguished, and thus our $\partial G$ is traditionally called `subdifferential'.} It is known that $G$ is differentiable at $x$ if and only if $\partial G(x)$ is a singleton, and if so, $\partial G(x)=\{DG(x)\}$.

If $n=1$, then define the left- and right-derivatives $D_-G(x),\ D_+G(x)$ such as
\[D_-G(x)=\lim_{y\uparrow x}\frac{G(y)-G(x)}{y-x},\ D_+G(x)=\lim_{y\downarrow x}\frac{G(y)-G(x)}{y-x}.\]
Note that, if $G$ is concave, then $\frac{G(y)-G(x)}{y-x}$ is nonincreasing in $y$, and thus
\[D_-G(x)=\inf_{t>0}\frac{G(x-t)-G(x)}{-t},\ D_+G(x)=\sup_{t>0}\frac{G(x+t)-G(x)}{t},\]
which implies that both $D_-G(x),D_+G(x)$ are defined and real numbers. It is known that $\partial G(x)=[D_+G(x),D_-G(x)]$.

Recall that, under Assumption 3, our $F$ is concave. In this case, the functions $c\mapsto F(k,c)$ and $k\mapsto F(k,c)$ are also concave, and thus the `partial' subdifferential can be considered. Let
\[\partial_kF(k,c)=\{p|F(k',c)-F(k,c)\le p(k'-k)\mbox{ for all }k'\ge 0\},\]
\[\partial_cF(k,c)=\{p|F(k,c')-F(k,c)\le p(c'-c)\mbox{ for all }c'\ge 0\}.\]
The partial left- and right-derivatives can be defined in the same manner. For example, 
\[D_{k,+}F(k,c)=\sup_{t>0}\frac{F(k+t,c)-F(k,c)}{t},\]
\[D_{k,-}F(k,c)=\inf_{t>0}\frac{F(k-t,c)-F(k,c)}{-t}.\]
We note one more fact. Suppose that $f$ is concave and continuous, $p\ge r\ge q, p\in \partial f(k_1), q\in \partial f(k_2)$ and $k_1<k_2$. We show that there exists $k\in [k_1,k_2]$ such that $r\in \partial f(k)$. If $r\ge D_+f(k_1)$, then $r\in \partial f(k_1)$. If $r\le D_-f(k_2)$, then $r\in \partial f(k_2)$. Therefore, we assume that $D_-f(k_2)<r<D_+f(k_1)$. Define $g(k)=f(k)-rk$. Then, $D_+g(k_1)>0$ and $D_-g(k_2)<0$, and thus, there exists $k\in ]k_1,k_2[$ such that $g(k)=\max_{k'\in [k_1,k_2]}g(k')$. By the definition of subdifferential, $0\in \partial g(k)$, and thus $r\in \partial f(k)$, as desired.

Applying this result for $k_1,k_2\ge 0$ with $k_1<k_2$, $c\ge 0$, and $g(k)=F(k,c)$, we obtain the {\bf mean value theorem} for a subdifferential. Indeed, we can easily verify that this function $g(k)$ is concave and $D_-g(k_1)\ge r,\ D_+g(k_2)\le r$ for $r=\frac{F(k_2,c)-F(k_1,c)}{k_2-k_1}$. Hence, there exists $k\in [k_1,k_2]$ such that $r\in \partial g(k)$. In summary, we have verified that, for every $k_1,k_2\ge 0$ with $k_1<k_2$ and $c\ge 0$, there exist $k\in [k_1,k_2]$ and $r\in \partial_kF(k,c)$ such that $F(k_2,c)-F(k_1,c)=r(k_2-k_1)$. We use this fact in the proof of Proposition \ref{Prop4}.

\subsection{Pure Accumulation Path}
Consider the following differential equation.
\begin{equation}
\dot{k}(t)=F(k(t),0),\ k(0)=\bar{k}.\label{PAP}
\end{equation}
Let $k^+(t,\bar{k})$ denote the solution to the above equation defined on $\mathbb{R}_+$. This function $k^+(t,\bar{k})$ is called the \textbf{pure accumulation path}. Later we prove the following lemma.

\begin{lem}\label{Lemma1}
Under Assumption 3, the pure accumulation path $k^+$ is uniquely defined on the set $\mathbb{R}_+\times\mathbb{R}_{++}$, and $\inf_{t\ge 0}k^+(t,\bar{k})>0$ for every $\bar{k}>0$.
\end{lem}

Let $V:\mathbb{R}_{++}\to \mathbb{R}$. The following requirement of $V$ is called the \textbf{growth condition}.
\begin{equation}
\lim_{T\to\infty}e^{-\rho T}V(k^+(T,\bar{k}))=0\mbox{ for all }\bar{k}>0.\label{GC}
\end{equation}
Define $\mathscr{V}$ as the space of all functions $V:\mathbb{R}_{++}\to \mathbb{R}$ that is increasing and concave, and satisfies the growth condition. Note that, by concavity, every $V\in \mathscr{V}$ is locally Lipschitz.

We should mention that (\ref{GC}) is not strong. Suppose that there exist $k>0$ and $\gamma\in \mathbb{R}$ such that $\gamma\in \partial_kF(k,0)$ and $\gamma<\rho$. Then, we can easily check that for all $\bar{k}>0$, there exist $A,B,C\in \mathbb{R}$ such that $k^+(t,\bar{k})\le Ae^{\gamma t}+Bt+C$.\footnote{See (\ref{SGC}) in the proof of Lemma \ref{Lemma1}.} Choose any increasing and concave function $V:\mathbb{R}_{++}\to \mathbb{R}$ and $p\in \partial V(\bar{k})$. Then,
\begin{align*}
-\infty<&~V(\inf_{t\ge 0}k^+(t,\bar{k}))\le V(k^+(T,\bar{k}))\\
\le&~V(\bar{k})+p(k^+(T,\bar{k})-\bar{k}),\\
\le&~V(\bar{k})+p(Ae^{\gamma T}+BT+C-\bar{k}),
\end{align*}
which implies that $V$ automatically satisfies (\ref{GC}). In this case, the requirement (\ref{GC}) vanishes and $\mathscr{V}$ coincides with the set of all increasing and concave functions on $\mathbb{R}_{++}$.

\vspace{12pt}
\noindent
{\bf A note on the growth condition}. We debated with many economists about what the growth condition means. One economist suggested that it might have something to do with the upper convergent condition, whereas another suggested that it might have something to do with the transversality condition. However, in our opinion, none of these conditions are the same as (\ref{GC}), just similar. We think that this condition is related to the terminal condition in the discrete-time optimization model. For example, equation (8) in Section 4.1 of Stokey and Lucas \cite{SL} is such a condition. This condition is crucial except for some cases in which this condition seems to be trivial, or some iteration techniques can be used such as Kamihigashi \cite{KA}. We think that terminal condition 1-(i) of Wiszniewska-Matyszkiel and Singh \cite{WMS} seems to be most similar in this context to (\ref{GC}). However, the terminal condition 1-(ii) of their paper seems to be independent of (\ref{GC}). Regarding why 1-(ii) is not necessary in this paper, we are not sure. It may be related to the fact that $k^*(t)$ in Proposition \ref{Prop4} satisfies $\inf_{t\ge 0}k^*(t)>0$.

\section{Results}
In this section, we analyze the HJB equation for a characterization of the value function. We prohibit ourselves from making assumptions on the solution explicitly. Assumptions must be made for properties of primitives $\rho, u, F$ in our model (\ref{MODEL}), and, for example, the assumption for the existence of the solution to (\ref{MODEL}) is not appropriate. The reason why we restrict ourselves is simple: in many cases, ensuring the existence of a solution to (\ref{MODEL}) is tremendously difficult. Hence, we prohibit ourselves from assuming the existence of a solution, even though under the existence assumption of the solution, the proofs of results become quite easy.

\subsection{Knowledge on Ordinary Differential Equations}
In this section, we frequently use knowledge on ODEs. Hence, we note basic knowledge on ODE for readers.

First, consider the following differential equation:
\begin{equation}
\dot{x}(t)=h(t,x(t)),\label{ODE}
\end{equation}
where $\dot{x}$ denotes $\frac{dx}{dt}$. We assume that $h:U\to \mathbb{R}^n$, $U\subset \mathbb{R}_+\times \mathbb{R}^n$, and the relative interior of $U$ in $\mathbb{R}_+\times \mathbb{R}^n$ is nonempty (denoted by $V$). We call a set $I\subset \mathbb{R}$ an \textbf{interval} if and only if $I$ is a convex set of $\mathbb{R}$ that includes at least two points. We say that a function $x:I\to \mathbb{R}^n$ is a \textbf{solution} to (\ref{ODE}) if and only if, 1) $I$ is an interval, 2) $x(t)$ is absolutely continuous on every compact subinterval of $I$, and 3) $\dot{x}(t)=h(t,x(t))$ for almost all $t\in I$. Suppose that $(t^*,x^*)\in U$. If a solution $x(t)$ to (\ref{ODE}) satisfies 1) $t^*\in I$ and 2) $x(t^*)=x^*$, then $x(t)$ is called a \textbf{solution with initial value condition} $x(t^*)=x^*$, or simply, a solution to the following differential equation:
\begin{equation}
\dot{x}(t)=h(t,x(t)),\ x(t^*)=x^*.\label{CODE}
\end{equation}
Now, suppose that $h:U\to \mathbb{R}^n$ satisfies the following requirements: 1) for every $t\in \mathbb{R}$, $x\mapsto h(t,x)$ is continuous, and 2) for every $x\in \mathbb{R}^n$, $t\mapsto h(t,x)$ is measurable. Then, we say that $h$ satisfies \textbf{Carath\'eodory's condition}. If, additionally, for a set $C\subset U$, there exists $L>0$ such that
\[\|h(t,x_1)-h(t,x_2)\|\le L\|x_1-x_2\|\]
for every $(t,x_1,x_2)$ such that $(t,x_1), (t,x_2)\in C$, then $h$ is called \textbf{Lipschitz in $x$ on $C$}.

The following facts are well known.
\begin{enumerate}[1)]
\item Suppose that $h$ satisfies Carath\'eodory's condition. Moreover, suppose that $(t^*,x^*)\in V$ and there exist $\varepsilon>0$ and an integrable function $r:[t^*-\varepsilon,t^*+\varepsilon]\to\mathbb{R}_+$ such that $\|h(t,x)\|\le r(t)$ for all $(t,x)\in U$ with $\|(t,x)-(t^*,x^*)\|\le \varepsilon$. Then, there exists a solution $x:I\to \mathbb{R}^n$ to (\ref{CODE}), where $I$ is relatively open in $\mathbb{R}_+$.\footnote{This result is called the Carath\'eodory-Peano existence theorem. For a proof, see Ch.2 of Coddington and Levinson \cite{CL1}.}

\item Suppose that $h$ satisfies Carath\'eodory's condition, and for every compact set $C\subset V$, $h$ is Lipschitz in $x$ on $C$. Moreover, suppose that $(t^*,x^*)\in V$ and there exists a convex neighborhood of $t^*$ such that $t\mapsto h(t,x^*)$ is integrable on this neighborhood. Then, there exists a solution $x:I\to \mathbb{R}^n$ to (\ref{CODE}), where $I$ is relatively open in $\mathbb{R}_+$.

\item Suppose that $h$ satisfies Carath\'eodory's condition, and for every compact set $C\subset V$, $h$ is Lipschitz in $x$ on $C$. Moreover, suppose that for every $(t^+,x^+)\in V$, there exists a convex neighborhood of $t^+$ such that $t\mapsto h(t,x^+)$ is integrable on this neighborhood. Choose any $(t^*,x^*)\in V$. Suppose that $x_1(t),x_2(t)$ are two solutions to (\ref{CODE}) such that $(t,x_i(t))\in V$ for every $t\in I_i$, where $I_i$ is the domain of $x_i(t)$. Then, $x_1(t)=x_2(t)$ for every $t\in I_1\cap I_2$.\footnote{Results 2) and 3) are known as the Carath\'eodory-Picard-Lindel\"of existence theorem. For a proof, see Section 0.4 of Ioffe and Tikhomirov \cite{IT}.}

\item Suppose that $h$ is continuous. Then, any solution $x(t)$ to (\ref{ODE}) is continuously differentiable.\footnote{See Ch.2 of Hartman \cite{HA}.}
\end{enumerate}

Next, suppose that $h$ satisfies all requirements in 3) and $(t^*,x^*)\in V$. Choose a solution $x:I\to \mathbb{R}^n$ to (\ref{CODE}). A solution $y:J\to \mathbb{R}^n$ is called an \textbf{extension} of $x$ if and only if 1) $I\subset J$, and 2) $y(t)=x(t)$ for all $t\in I$. Then, $x(t)$ is called \textbf{nonextendable} if and only if there is no extension except $x(t)$ itself.

The following facts are well known.\footnote{Fact 5) can be proved easily. The proof of 6) is in Ch.2 of Coddington and Levinson \cite{CL1}.}
\begin{enumerate}[1)]
\setcounter{enumi}{4}
\item In addition to the requirements of 3), suppose that $V=U$. Then, there uniquely exists a nonextendable solution $x(t)$ to (\ref{CODE}). Moreover, the domain $I$ of $x(t)$ is relatively open in $\mathbb{R}_+$.

\item Suppose that all requirements of 5) hold, and let $x:I\to \mathbb{R}^n$ be the nonextendable solution to (\ref{CODE}). Choose any compact set $C\subset V$. Then, there exists $t^+\in I$ such that if $t^+\le t\in I$, then $(t,x(t))\notin C$.
\end{enumerate}
Finally, suppose that $h(t,x)=a(t)x+b(t)$, where $a(t), b(t)$ are locally integrable functions defined on $\mathbb{R}_+$ and $a(t)$ is bounded. Then, the solution to (\ref{ODE}) is determined by the following formula:
\begin{equation}
x(t)=e^{\int_0^ta(\tau)d\tau}\left[x(0)+\int_0^te^{-\int_0^sa(\tau)d\tau}b(s)ds\right].\label{LODE}
\end{equation}
This is called the formula of the solution for linear ODEs.

\subsection{Lemmas and Their Proofs}
We should prove Lemma \ref{Lemma1}. However, to derive Lemma \ref{Lemma1}, we must introduce two more results on differential equations. Because these are also useful for later arguments, we provide these results as lemmas.

\begin{lem}\label{Lemma2}
Consider the following two ODEs:
\begin{equation}
\dot{k}(t)=h_i(t,k),\ k(0)=\bar{k}_i,\label{eq:eq13}
\end{equation}
where $i\in \{1,2\}$. Suppose that each $h_i$ is a real valued function defined on some convex neighborhood $U\subset \mathbb{R}_+\times\mathbb{R}$ of $(0,\bar{k}_i)$ and $h_i$ satisfies Carath\'eodory's condition. Then, the following results hold.
\begin{enumerate}[i)]
\item For some $i\in \{1,2\}$, if there exists a locally integrable function $r(t)$ such that
\[\sup_{k:(t,k)\in U}|h_i(t,k)|\le r(t),\]
then there exists $T>0$ such that this equation $(\ref{eq:eq13})$ has a solution $k_i:[0,T]\to\mathbb{R}$. Moreover, if $h_i(t,k)$ is continuous, then $k_i(t)$ is continuously differentiable.

\item Suppose that $\bar{k}_1\le \bar{k}_2$, $h_1(t,k)\le h_2(t,k)$ for every $(t,k)\in U$, and for some $i^*\in \{1,2\}$, there exists $L>0$ such that if $(t,k_1),(t,k_2)\in U$, then
\[|h_{i^*}(t,k_1)-h_{i^*}(t,k_2)|\le L|k_1-k_2|.\]
Suppose also that $h_{i^*}(t,\bar{k})$ is locally integrable, and there exist solutions $k_i:[0,T]\to \mathbb{R}$ to the above equations for $i\in \{1,2\}$. Then, $k_1(t)\le k_2(t)$ for all $t\in [0,T]$.\footnote{If such an $L>0$ is absent, then this lemma does not hold. For example, consider $\bar{k}_1=\bar{k}_2=0,\ h_1(t,k)=\sqrt{|k|}-\frac{t}{8},\ h_2(t,k)=\sqrt{|k|},\ k_1(t)=\frac{t^2}{16},$ and $k_2(t)\equiv 0$.

Note that if $\bar{k}_1=\bar{k}_2$ and $h_1=h_2$, then this claim immediately implies the uniqueness of the solution. For this view, this lemma is an extension of the Carath\'eodory-Picard-Lindel\"of uniqueness result in the theory of ODEs.}
\end{enumerate}
\end{lem}

\begin{proof}
Assertion i) is just a corollary of the Carath\'eodory-Peano existence theorem, and thus we omit the proof.

Suppose that assertion ii) does not hold. Then, there exists $t^*>0$ such that $k_1(t^*)>k_2(t^*)$. Let $t^+=\inf\{t\in [0,t^*]|k_1(s)>k_2(s)\mbox{ for all }s\in [t,t^*]\}$. Because $k_1(0)=\bar{k}_1\le \bar{k}_2=k_2(0)$, we have that $k_1(t^+)=k_2(t^+)$. We treat only the case $i^*=2$, because the case $i^*=1$ can be treated symmetrically. Define
\[k_3(t)=k_1(t^+)+\int_{t^+}^th_2(s,k_1(s))ds.\]
Because $h_2$ satisfies Carath\'eodory's condition, the mapping $t\mapsto h_2(t,k_1(t))$ is measurable,\footnote{See Section 8.1 of Ioffe and Tikhomirov \cite{IT}.} and
\[|h_2(t,k_1(t))|\le |h_2(t,k_1(t))-h_2(t,\bar{k})|+|h_2(t,\bar{k})|\le L|k_1(t)-\bar{k}|+|h_2(t,\bar{k})|,\]
and thus $h_2(t,k_1(t))$ is locally integrable, which implies that $k_3(t)$ is defined on $[t^+,t^*]$. Moreover,
\[k_3(t)\ge k_1(t^+)+\int_{t^+}^th_1(s,k_1(s))ds=k_1(t)\ge k_2(t)\]
for every $t\in [t^+,t^*]$. On the other hand,
\begin{align*}
k_3(t)-k_2(t)=&~\int_{t^+}^t[h_2(s,k_1(s))-h_2(s,k_2(s))]ds\\
\le&~\int_{t^+}^tL[k_1(s)-k_2(s)]ds\\
\le&~L(t-t^+)\max_{s\in [t^+,t]}(k_1(s)-k_2(s)).
\end{align*}
Fix some $t\in [t^+,t^*]$ with $0<t-t^+<L^{-1}$, and let $s^*\in \arg\max\{k_1(s)-k_2(s)|s\in [t^+,t]\}$. Because $k_1(t^+)=k_2(t^+)$ and $k_1(s)>k_2(s)$ for $s\in ]t^+,t^*]$, $s^*>t^+$. Then,
\[k_3(s^*)-k_2(s^*)\le L(s^*-t^+)(k_1(s^*)-k_2(s^*))<k_1(s^*)-k_2(s^*),\]
and thus $k_3(s^*)<k_1(s^*)$, which is a contradiction.
\end{proof}

\begin{lem}\label{Lemma3}
Consider the following ODE:
\begin{equation}
\dot{k}(t)=h(t,k(t)),\ k(0)=\bar{k}>0, \label{eq:eq14}
\end{equation}
where $h:\mathbb{R}_+\times\mathbb{R}_{++}\to\mathbb{R}$ satisfies Carath\'eodory's condition, and for every $(t^+,k^+)\in \mathbb{R}_+\times\mathbb{R}_{++}$, there exist $\varepsilon>0$ and a positive and locally integrable function $r(t)$ defined on a neighborhood of $t^+$ such that if $|k-k^+|\le \varepsilon$, then $|h(t,k)|<r(t)$ for every $t$ such that $r(t)$ is defined. For either $X=\mathbb{R}_+$ or $X=[0,T^+]$ with $T^+>0$, suppose that there exist two positive continuous functions $\hat{k}(t)$ and $\bar{k}(t)$ defined on $X$ such that $\bar{k}(t)\le k(t)\le \hat{k}(t)$ for any solution $k(t)$ to $(\ref{eq:eq14})$ defined on $[0,T]$ for some $T>0$ and $t\in [0,T]\cap X$. Then, this equation $(\ref{eq:eq14})$ has a solution defined on $X$ itself.
\end{lem}

\begin{proof}
We treat only the case $X=\mathbb{R}_+$. The proof in the case $X=[0,T^+]$ is almost the same.

By Lemma \ref{Lemma2}, there exists $T>0$ such that equation (\ref{eq:eq14}) has a solution $k(t)$ defined on $[0,T]$. By assumption, $\bar{k}(t)\le k(t)\le \hat{k}(t)$ for all $t\in [0,T]$. 

Next, let $Y$ be the set of all solutions to equation (\ref{eq:eq14}) defined on either $\mathbb{R}_+$ or $[0,T']$ for some $T'>0$, and let $k_1(\cdot)\succeq k_2(\cdot)$ if the domain of $k_1(\cdot)$ includes that of $k_2(\cdot)$ and $k_1(t)=k_2(t)$ when both are defined at $t$. Clearly, $\succeq$ is a partial order on $Y$. For $k(\cdot)\in Y$, let $I_{k(\cdot)}$ be the domain of $k(\cdot)$. Choose any chain $C\subset Y$ of $\succeq$. If $\sup \cup_{k(\cdot)\in C}I_{k(\cdot)}=+\infty$, then we can define
\[k^+(t)=k(t)\mbox{ if }t\in I_{k(\cdot)},\]
and $k^+(\cdot)$ is an upper bound of $C$. Otherwise, let $T^*=\sup \cup_{k(\cdot)\in C}I_{k(\cdot)}$. Define
\[k^+(t)=k(t)\mbox{ if }t\in I_{k(\cdot)}.\]
Then, $k^+(t)$ is a solution to (\ref{eq:eq14}) defined on $[0,T^*[$. By the continuity of $\hat{k}(t),\bar{k}(t)$, there exist $\varepsilon>0$ and $\delta>0$ such that $\bar{k}(T^*)>\varepsilon$ and if $0<T^*-t<\delta$, then $k^+(t)\in [\bar{k}(T^*)-\varepsilon, \hat{k}(T^*)+\varepsilon]$. Hence, we can define
\[k^+(T^*)=\limsup_{t\to T^*}k^+(t)\in [\bar{k}(T^*)-\varepsilon,\hat{k}(T^*)+\varepsilon].\]
Because $[\bar{k}(T^*)-\varepsilon, \hat{k}(T^*)+\varepsilon]$ is compact, if $\delta>0$ is sufficiently small, then there exists a positive and locally integrable function $r(t)$ such that $|h(t,k)|<r(t)$ for every $(t,k)$ such that $0<T^*-t<\delta$ and $k\in [\bar{k}(T^*)-\varepsilon,\hat{k}(T^*)+\varepsilon]$. Thus, $|k^+(t)-k^+(T^*)|\le \int_t^{T^*}r(s)ds$, and hence
\[\lim_{t\to T^*}k^+(t)=k^+(T^*),\]
which implies that $k^+(t)$ is continuous on $[0,T^*]$. Because
\[k^+(t)=\bar{k}+\int_0^th(s,k^+(s))ds\]
for all $t\in [0,T^*]$, we conclude that $k^+(t)$ is a solution to equation (\ref{eq:eq14}) defined on $[0,T^*]$, and thus $k^+(\cdot)\in Y$ and it is an upper bound of $C$. Therefore, by Zorn's lemma, there exists a maximal element $k^*(\cdot)\in Y$ of $\succeq$. If the domain $I_{k^*(\cdot)}=[0,T]$, then $k^*(T)\in [\bar{k}(T),\hat{k}(T)]$, and thus by Lemma \ref{Lemma2}, there exists a solution $\tilde{k}(t)$ to the following equation:
\[\dot{k}(t)=h(t+T,k(t)),\ k(0)=k^*(T),\]
defined on $[0,T']$. Define
\[\kappa(t)=\begin{cases}
k^*(t) & \mbox{if }t\in [0,T],\\
\tilde{k}(t-T) & \mbox{if }t\in [T,T+T'].
\end{cases}\]
Then $\kappa(\cdot)\in Y$ and $\kappa(\cdot)\succ k^*(\cdot)$, which is a contradiction. Therefore, the domain of $k^*(t)$ must be $\mathbb{R}_+$. This completes the proof of Lemma \ref{Lemma3}.
\end{proof}

\begin{proof}[{\bf Proof of Lemma \ref{Lemma1}}]
Suppose that Assumption 3 holds, and consider the following ODE:
\[\dot{k}(t)=F(k(t),0),\ k(0)=\bar{k}.\]
By Assumption 3, there exists $k>0$ such that $F(k,0)>0$. Suppose that a solution $k(t)$ to the above equation satisfies $k(T)<\min\{\bar{k},k\}$ for some $T>0$, and define $t^*=\inf\{t\in [0,T]|\forall t'\in [t,T], k(t')<\min\{\bar{k},k\}\}$. Because $k(0)=\bar{k}\ge \min\{\bar{k},k\}$, $k(t^*)=\min\{\bar{k},k\}$. Because $F$ is concave and $F(0,0)=0$, $F(k(t^*),0)>0$. Thus, there exists $t\in ]t^*,T[$ such that $k(t)>\min\{\bar{k},k\}$, which is a contradiction. Therefore,
\[k(t)\ge \min\{\bar{k},k\}\]
for every solution $k(t)$ to the above equation. Meanwhile, let $p\in \partial_kF(k,0)$. Then, $F(k',0)\le p(k'-k)+F(k,0)$, and thus, by Lemma \ref{Lemma2} and the formula (\ref{LODE}),
\begin{equation}
k(t)\le \begin{cases}
e^{pt}\left(\bar{k}+(e^{-pt}-1)\frac{pk-F(k,0)}{p}\right) & \mbox{if }p\neq 0,\\
\bar{k}+tF(k,0) & \mbox{if }p=0,
\end{cases}\label{SGC}
\end{equation}
for every solution $k(t)$ to the above equation. Therefore, we can apply Lemma \ref{Lemma3}, and thus $k^+(t,\bar{k})$ is well-defined on $\mathbb{R}_+\times\mathbb{R}_{++}$, and $\inf_{t\ge 0}k^+(t,\bar{k})\ge \min\{\bar{k},k\}>0$ for every $\bar{k}>0$. This completes the proof of Lemma \ref{Lemma1}.
\end{proof}

Note that, if $h(t,k)=F(k,g(t,k))$ for some non-negative function $g(t,k)$, then we can use $k^+(t,\bar{k})$ for $\hat{k}(t)$ in Lemma \ref{Lemma3}. Suppose that $g(t,k)=c(t)$ for some $c(t)\in W$. Then, for all $k\in [0,\bar{k}+1]$,\footnote{Recall that, by Assumption 3, if $W=W_1$, then $\delta_2(c)=d_2c$.}
\begin{equation}
|h(t,k)|\le d_1(\bar{k}+1)+\delta_2(c(t))+\max\{F(k',0)|0\le k'\le \bar{k}+1\}\equiv r(t),\label{RT}
\end{equation}
and because $r(t)$ is locally integrable, the requirement of Lemma \ref{Lemma3} is automatically satisfied. We use these facts frequently.

Later, we will often need the boundedness of $c(t)$ in the proof. Therefore, we present two more lemmas.\footnote{We assume that if $W=W_1$, then $\delta_2(c)=d_2c$. Actually, the reason why such an assumption is needed is in the proof of Lemma \ref{Lemma5}. If $W=W_1$ and $\delta_2(c)\neq d_2c$, then $r(t)$ in the proof of Lemma \ref{Lemma5} may not be locally integrable, and our proof is broken.}

\begin{lem}\label{Lemma4}
Suppose that Assumption 3 holds, and choose any nonnegative, measurable, and bounded function $c(t)$. Define
\[h(t,k)=F(k,c(t)).\]
Then, $h$ satisfies Carath\'eodory's condition on $\mathbb{R}_+^2$ and for every compact set $C\subset \mathbb{R}_+\times \mathbb{R}_{++}$, it is Lipschitz in $k$ on $C$.
\end{lem}

\begin{proof}
The fact that $h$ satisfies Carath\'eodory's condition can easily be shown. Choose any compact set $C\subset \mathbb{R}_+\times \mathbb{R}_{++}$. Then, there exist $T>0$ and $\tilde{k},\hat{k}>0$ such that $C\subset [0,T]\times [\tilde{k},\hat{k}]$. Define $c^*=\sup_{t\ge 0}c(t)$ and
\[\tilde{F}(k,c)=\begin{cases}
F(k,c) & \mbox{if }c\ge 0,\\
F(k,0) & \mbox{if }c<0.
\end{cases}\]
Then, $\tilde{F}$ is defined on $\mathbb{R}_{++}\times \mathbb{R}$. We can easily check that $\tilde{F}$ is concave. By Theorem 24.7 of Rockafeller \cite{ROC}, the set
\[X=\{(p,q)|(p,q)\in \partial \tilde{F}(k,c)\mbox{ for }(k,c)\in [\tilde{k},\hat{k}]\times [0,c^*]\}\]
is compact.\footnote{Note that $[\tilde{k},\hat{k}]\times [0,c]$ is in the relative interior of the domain of $\tilde{F}$, and thus Theorem 24.7 of Rockafeller \cite{ROC} is applicable for $\tilde{F}$.} Thus, for $t\ge 0$ and $k_1,k_2\in [\tilde{k},\hat{k}]$,
\begin{align*}
|h(t,k_1)-h(t,k_2)|=&~|\tilde{F}(k_1,c(t))-\tilde{F}(k_2,c(t))|\\
\le&~\max\{|p||(p,q)\in X\}\times |k_1-k_2|,
\end{align*}
as desired. This completes the proof.
\end{proof}

\begin{lem}\label{Lemma5}
Suppose that Assumptions 1-3 hold. Choose any $\bar{k}>0$ and $(k^*(t),c^*(t))\in A_{\bar{k}}$ such that $\int_0^{\infty}e^{-\rho t}u(c^*(t),k^*(t))dt>-\infty$, and define $c_n(t)=\min\{c^*(t),n\}$ for $n\in \mathbb{N}$. Then, there exists a function $k_n(t)$ defined on $\mathbb{R}_+$ such that $k_n(t)\ge k^*(t)$ for all $t$, and if $u(n,0)>-\infty$, then $(k_n(t),c_n(t))\in A_{\bar{k}}$ and $\int_0^{\infty}e^{-\rho t}u(c_n(t),k_n(t))dt>-\infty$.
\end{lem}

\begin{proof}
First, define $h(t,k)=F(k,c^*(t))$ and $h_n(t,k)=F(k,c_n(t))$. Then, $h_n(t,k)\ge h(t,k)$ for all $t\ge 0$. Consider the following differential equation:
\[\dot{k}(t)=h_n(t,k(t)),\ k(0)=\bar{k}.\]
We show that there exists a solution $k_n(t)$ to the above equation defined on $\mathbb{R}_+$ such that $k_n(t)\ge k^*(t)$ for every $t\ge 0$.

Let $Y$ be the set of all solutions $k(\cdot)$ to the above equation such that the domain $I_{k(\cdot)}$ is either $\mathbb{R}_+$ or $[0,T]$ for some $T>0$, and $k(t)\ge k^*(t)$ for all $t\in I_{k(\cdot)}$. By Lemma \ref{Lemma4}, $h_n$ satisfies all requirements in Lemma \ref{Lemma2}, and thus $Y$ is nonempty.

Define a partial order $\succeq$ on $Y$ such that $k_1(\cdot)\succeq k_2(\cdot)$ if and only if $I_{k_2(\cdot)}\subset I_{k_1(\cdot)}$ and $k_1(t)=k_2(t)$ for every $t\in I_{k_2(\cdot)}$. Let $C$ be any chain of $\succeq$. If $\sup\cup_{k(\cdot)\in C}I_{k(\cdot)}=+\infty$, we can define $k^+(t)=k(t)$ for $t\in I_{k(\cdot)}$, and $k^+(\cdot)$ is an upper bound of $C$. If $\sup\cup_{k(\cdot)\in C}I_{k(\cdot)}=T<+\infty$, we can define $k^+(t)=k(t)$ for $t\in I_{k(\cdot)}$, and $k^+(t)$ is defined on $[0,T[$. Because $k^+(t)\le k^+(t,\bar{k})$, $k^+(T)=\limsup_{t\to T}k^+(t)\in [0,k^+(T,\bar{k})]$ can be defined. Let $r(t)$ be defined as in (\ref{RT}), where we use $\max_{t\in [0,T]}k^+(t,\bar{k})$ instead of $\bar{k}$ and $c^*(t)$ instead of $c(t)$.\footnote{That is,
\[r(t)=d_1(\max_{t\in [0,T]}k^+(t,\bar{k})+1)+\delta_2(c^*(t))+\max\{F(k,0)|0\le k\le \max_{t\in [0,T]}k^+(t,\bar{k})+1\}.\]
We frequently use such a construction of $r(t)$ later.} Then, $|k^+(t)-k^+(T)|\le \int_t^Tr(s)ds$, and thus $k^+(\cdot)$ is continuous on $[0,T]$. Because
\[k^+(t)=\bar{k}+\int_0^th_n(s,k^+(s))dt\]
for every $t\in [0,T]$, $k^+(\cdot)$ is a solution to the above equation. Because both $k^+(\cdot),k^*(\cdot)$ are continuous, $k^+(t)\ge k^*(t)$ for every $t\in [0,T]$, and thus $k^+(\cdot)$ is an upper bound of $C$. Therefore, by Zorn's lemma, there exists a maximal element $\tilde{k}(\cdot)$. Suppose that $I_{\tilde{k}(\cdot)}=[0,T]\neq \mathbb{R}_+$. If $\tilde{k}(T)>0$, then using Lemma \ref{Lemma2}, we can easily show that $\tilde{k}(\cdot)$ is not maximal, which is a contradiction. Therefore, $\tilde{k}(T)=k^*(T)=0$. Choose any $k^*>0$ such that $k\mapsto F(k,c)$ is increasing on $[0,k^*]$ for all $c\in [0,n]$. Define $r(t)$ as in (\ref{RT}), where we use $k^*$ instead of $\bar{k}$ and $c^*(t)$ instead of $c(t)$. Choose $\varepsilon>0$ so small that $k^*(t)\le k^*$ whenever $t\in [T,T+\varepsilon]$, and
\[\int_T^{T+\varepsilon}r(t)dt<k^*.\]
Let $I=[T,T+\varepsilon]$ and $X$ be the space of all continuous functions $x:I\to [0,k^*]$ such that $x(T)=0$, $x(t)\ge k^*(t)$ for all $t\in I$, and for every $t_1,t_2\in I$ with $t_1<t_2$, $|x(t_1)-x(t_2)|\le \int_{t_1}^{t_2}r(t)dt$. Because $k^*(\cdot)$ satisfies all requirements, $X$ is nonempty. Clearly, $X$ is convex. By Ascoli-Arzela's theorem, $X$ is compact with respect to the sup norm. Define
\[P(x(\cdot))(t)=\int_T^th_n(s,x(s))ds.\]
If $x(\cdot)\in X$ and $y(\cdot)=P(x(\cdot))$, then
\[y(T)=0,\]
\[y(t)\ge \int_T^th_n(s,k^*(s))ds\ge \int_T^th(s,k^*(s))ds=k^*(t),\]
and if $t_1,t_2\in I$ and $t_1<t_2$, then
\[|y(t_1)-y(t_2)|\le \int_{t_1}^{t_2}|h_n(t,x(t))|dt\le \int_{t_1}^{t_2}r(t)dt.\]
Therefore,
\[y(t)\le \int_T^{T+\varepsilon}r(t)dt<k^*,\]
and thus $y(\cdot)\in X$, which implies that $P$ is a mapping from $X$ into $X$. Let $(x_m(\cdot))$ be a convergent sequence of $X$ on the sup norm and $x(\cdot)$ is the limit. Define $y_m(\cdot)=P(x_m(\cdot))$ and $y(\cdot)=P(x(\cdot))$. Choose any $\varepsilon'>0$ and define
\[B_m=\left\{t\in I\left||h_n(t,x_m(t))-h_n(t,x(t))|\ge \frac{\varepsilon'}{2\varepsilon}\right.\right\},\]
\[C_m=\cup_{\ell\ge m}B_{\ell}.\]
Because $F$ is continuous, $C_m\downarrow \emptyset$ as $m\to \infty$, and thus
\[\lim_{m\to \infty}\int_{C_m}r(t)dt=0.\]
Hence, there exists $m_0\in \mathbb{N}$ such that if $m\ge m_0$, then
\[\int_{C_m}2r(t)dt<\frac{\varepsilon'}{2}.\]
Then, for every $t\in I$, if $m\ge m_0$,
\begin{align*}
|y_m(t)-y(t)|\le&~\int_T^{T+\varepsilon}|h_n(\tau,x_m(\tau))-h_n(\tau,x(\tau))|d\tau\\
=&~\int_{I\setminus C_m}|h_n(\tau,x_m(\tau))-h_n(\tau,x(\tau))|d\tau\\
&~+\int_{C_m}|h_n(\tau,x_m(\tau))-h_n(\tau,x(\tau))|d\tau\\
\le&~\int_{I\setminus C_m}\frac{\varepsilon'}{2\varepsilon}d\tau+\int_{C_m}2r(\tau)d\tau\\
<&~\frac{\varepsilon'}{2}+\frac{\varepsilon'}{2}=\varepsilon',
\end{align*}
which implies that $P$ is continuous with respect to the sup norm. Therefore, by Schauder's fixed point theorem, there exists a fixed point $x^*(\cdot)\in X$ of $P$. Define
\[\hat{k}(t)=\begin{cases}
\tilde{k}(t) & \mbox{if }0\le t\le T,\\
x^*(t) & \mbox{if }t\in I.
\end{cases}\]
Then, $\hat{k}(\cdot)\in Y$ and $\hat{k}(\cdot)\succ \tilde{k}(\cdot)$, which contradicts the maximality of $\tilde{k}(\cdot)$. Therefore, the domain of $\tilde{k}(\cdot)$ is $\mathbb{R}_+$, and thus we can define $k_n(t)=\tilde{k}(t)$.

Now, suppose that $u(n,0)>-\infty$. Because $\int_0^{\infty}e^{-\rho t}u(c^*(t),k^*(t))dt>-\infty$, for $u^-(c,k)=\min\{u(c,k),0\}$,
\[\int_{\{t|c^*(t)\le n\}}e^{-\rho t}u^-(c^*(t),k^*(t))dt>-\infty,\]
and thus,
\begin{align*}
-\infty<&~\int_{\{t|c^*(t)\le n\}}e^{-\rho t}u^-(c^*(t),k^*(t))dt+\int_{\{t|c^*(t)>n\}}e^{-\rho t}u^-(n,k^*(t))dt\\
=&~\int_0^{\infty}e^{-\rho t}u^-(c_n(t),k^*(t))dt\le\int_0^{\infty}e^{-\rho t}u^-(c_n(t),k_n(t))dt.
\end{align*}
Hence, $\int_0^{\infty}e^{-\rho t}u(c_n(t),k_n(t))dt$ is defined and is not $-\infty$, which implies that $(k_n(t),c_n(t))\in A_{\bar{k}}$. This completes the proof.
\end{proof}

Note that, if $(k^*(t),c^*(t)), (k_n(t),c_n(t))\in A_{\bar{k}}$ are as in Lemma \ref{Lemma5}, then $\int_0^{\infty}e^{-\rho t}u(c_n(t),k^*(t))dt>-\infty$ for sufficiently large $n$. By the monotone convergence theorem,
\begin{equation}\label{FIN}
\liminf_{n\to \infty}\int_0^{\infty}e^{-\rho t}u(c_n(t),k_n(t))dt\ge \int_0^{\infty}e^{-\rho t}u(c^*(t),k^*(t))dt.
\end{equation}
In particular, if there exists $(k^*(t),c^*(t))\in A_{\bar{k}}$ such that $\int_0^{\infty}e^{-\rho t}u(c^*(t),k^*(t))dt>M$, then we can assume that $c^*(t)$ is bounded.

\subsection{Basic Propositions}
In this subsection, we treat two propositions that can be proved using only Assumptions 1-4.

\begin{prop}\label{Prop1}
 Suppose that Assumptions 1-4 hold. Then, there exists a positive continuous function $c^*(p,k)$ defined on $\mathbb{R}^2_{++}$ such that for all $(p,k)\in \mathbb{R}^2_{++}$,
\[F(k,c^*(p,k))p+u(c^*(p,k),k)=\sup_{c\ge 0}\{F(k,c)p+u(c,k)\}.\]
\end{prop}

\begin{proof}
Let
\[g(c,k,p)=F(k,c)p+u(c,k).\]
Because $\frac{\partial u}{\partial c}$ is decreasing in $c$, $g$ is strictly concave with respect to $c$. Fix $p,k>0$. Choose $\varepsilon>0$ so small that $p-\varepsilon>0$ and $k-\varepsilon>0$. Define
\[\varepsilon^*=\min_{k'\in [k-\varepsilon, k+\varepsilon]}[F(k,0)-F(k,1)].\]
Then, for every $k'\in [k-\varepsilon, k+\varepsilon],\ q\in [p-\varepsilon,p+\varepsilon]$ and $c\ge 1$,
\begin{equation}
D_{c,-}g(c,k',q)\le \max_{k''\in [k-\varepsilon, k+\varepsilon]}\frac{\partial u}{\partial c}(c,k'')-(p-\varepsilon)\varepsilon^*.\label{EQ}
\end{equation}
Because $\frac{\partial u}{\partial c}(c,k'')\to 0$ as $c\to \infty$, there exists $\bar{c}>1$ such that the right-hand side of (\ref{EQ}) is negative for all $c\ge \bar{c}$.\footnote{Use Dini's theorem.} Then, for such $k'$ and $q$, $g(c,k',q)<g(\bar{c},k',q)$ for all $c>\bar{c}$, and thus,
\[\sup_{c\ge 0}g(c,k',q)=\max_{0\le c\le \bar{c}}g(c,k',q).\]
The remainder of the proof on the continuity of the function $c^*$ at $(p,k)$ is a simple application of Berge's theorem. The positivity of $c^*(p,k)$ follows from the assumption $\lim_{c\to 0}\frac{\partial u}{\partial c}(c,k)=+\infty$.
\end{proof}

Note that, if $F(k,c)=f(k)-c$ and $u(c,k)=u(c)$, then by the first-order condition, $c^*(p,k)=(u')^{-1}(p)$, and thus $c^*(p,k)$ is independent of $k$. This result can be extended. If $F(k,c)=f(k)-h(c)$, then by the first-order condition, $c^*(p,k)$ is the unique $c$ such that $h'(c)p=u'(c)$, and thus $c^*(p,k)$ is also independent of $k$.

\begin{prop}\label{Prop2}
Suppose that Assumptions 1-3 hold. Then, $\bar{V}(\bar{k})>-\infty$ for every $\bar{k}>0$ and the function $\bar{V}$ is nondecreasing and concave.
\end{prop}

By Proposition \ref{Prop2}, $\bar{V}(k)\in\mathbb{R}$ for some $k>0$ if and only if $\bar{V}(k)\in \mathbb{R}$ for every $k>0$. We say that $\bar{V}$ is \textbf{finite} if $\bar{V}(k)\in\mathbb{R}$ for every $k>0$.

\begin{proof}
We separate the proof into three steps.

\begin{step}
$\bar{V}(\bar{k})>-\infty$ for every $\bar{k}>0$.
\end{step}

\begin{proof}[{\bf Proof of Step 1}]
We have assumed that $F(0,0)=0$ and there exists $k^*>0$ such that $F(k^*,0)>0$. If $F(\bar{k},0)>0$, then there exists $\bar{c}>0$ such that $F(\bar{k},\bar{c})=0$ because $F$ is continuous, concave, and decreasing in $c$.\footnote{By this assumption, $\lim_{c\to \infty}F(\bar{k},c)=-\infty$.} Hence, $(k(t),c(t))\equiv (\bar{k},\bar{c})\in A_{\bar{k}}$, and thus $\bar{V}(\bar{k})>-\infty$. If $F(\bar{k},0)\le 0$, then by the concavity of $F$, $\bar{k}>k^*$. Choose any $c^*>0$ such that $F(k^*,c^*)>0$, and any solution $k(t)$ to the following differential equation
\[\dot{k}(t)=F(k(t),c^*),\ k(0)=\bar{k},\]
defined on $[0,T]$. Suppose that $k(t^+)<k^*$ for some $t^+>0$. Let $t^*=\inf\{t\in [0,t^+]|\forall s\in [t,t^+], k(s)<k^*\}$. Because $k(0)=\bar{k}>k^*$, we have that $k(t^*)=k^*$, and thus $F(k(t^*),c^*)>0$. Therefore, there exists $t\in ]t^*,t^+]$ such that $k(t)>k^*$, which is a contradiction. Hence, $k(t)\ge k^*$ for all $t\in [0,T]$, and thus by Lemma \ref{Lemma3}, there exists a solution $k^*(t)$ to the above equation defined on $\mathbb{R}_+$. Let $c^*(t)\equiv c^*$. Then, $(k^*(t),c^*(t))\in A_{\bar{k}}$, and thus $\bar{V}(\bar{k})>-\infty$. This completes the proof of Step 1.
\end{proof}

\begin{step}
$\bar{V}$ is nondecreasing.\footnote{The conclusions of Steps 2-3 still hold even if we define $\bar{V}(0)$ and consider that the domain of $\bar{V}$ is $\mathbb{R}_+$. The proof is easy.}
\end{step}

\begin{proof}[{\bf Proof of Step 2}]
Choose any $k_1,k_2>0$ with $k_1<k_2$. In Step 1, we have proved that there exists a pair $(k(t),c(t))\in A_{k_1}$, and $\bar{V}(k_1)>-\infty$. Let $(k(t),c(t))\in A_{k_1}$ and
\[\int_0^{\infty}e^{-\rho t}u(c(t),k(t))dt>\min\{\bar{V}(k_1)-n^{-1},n\}.\]
Because of Lemma \ref{Lemma5}, we can assume that $c(t)$ is bounded. Define
\[h(t,k)=F(k,c(t))\]
for all $(t,k)\in\mathbb{R}_+\times\mathbb{R}_{++}$, and consider the following two ODEs:\footnote{Note that, equation (\ref{ODEH2}) is different from equation (\ref{ODEH}), because the right-hand side of (\ref{ODEH}) is defined only on $\mathbb{R}_+\times\mathbb{R}_{++}$, while that of (\ref{ODEH2}) is defined on $\mathbb{R}^2_+$. The reason why we need (\ref{ODEH}) is as follows: the equation (\ref{ODEH2}) does not satisfy the requirement $U=V$ in 5) of Subsection 4.1, and thus 5) cannot be directly applied.}
\begin{align}
\dot{\kappa}(t)=&~h(t,\kappa(t)),\ \kappa(0)=k_2,\label{ODEH}\\
\dot{\kappa}(t)=&~F(\kappa(t),c(t)),\ \kappa(0)=k_2.\label{ODEH2}
\end{align}
Suppose that there does not exist a solution $\kappa:\mathbb{R}_+\to \mathbb{R}_+$ to (\ref{ODEH2}) such that $\kappa(t)\ge k(t)$ for every $t\in\mathbb{R}_+$. Because all assumptions in 5) of Subsection 4.1 hold for equation (\ref{ODEH}), there exists the unique nonextendable solution $\kappa:I\to \mathbb{R}$ to (\ref{ODEH}). By our assumption, $\sup\{t\in I|\kappa(t)\ge k(t)\}=T<+\infty$. If $[0,T]\subset I$, then $\kappa(T)=k(T)$, and thus if we define
\[\tilde{k}(t)=\begin{cases}
\kappa(t) & \mbox{if }0\le t<T,\\
k(t) & \mbox{otherwise},
\end{cases}\]
then $\tilde{k}(t)$ is a solution to (\ref{ODEH2}) defined on $\mathbb{R}_+$ such that $\tilde{k}(t)\ge k(t)$ for all $t\in \mathbb{R}_+$, which is a contradiction. Therefore, we must have that $I=[0,T[$. Because $\kappa(t)\le k^+(t,k_2)$ for all $t\in I$, by 6) in Subsection 4.1, for every $n\in \mathbb{N}$, there exists $t_n>0$ such that if $t_n<t<T$, then $\kappa(t)<n^{-1}$. Therefore, $\limsup_{t\to T}\kappa(t)=0$, and thus we can define $\kappa(T)=0$. Because $\kappa(t)\ge k(t)$ for all $t\in I$, $k(T)=0$, and thus we can define $\tilde{k}(t)$ as above, which leads to a contradiction.

Therefore, there exists a solution $\kappa(t)$ to (\ref{ODEH2}) defined on $\mathbb{R}_+$ such that $\kappa(t)\ge k(t)$ for every $t\ge 0$, which implies that $(\kappa(t),c(t))\in A_{k_2}$. Hence,
\[\bar{V}(k_2)\ge \int_0^{\infty}e^{-\rho t}u(c(t),k(t))dt,\]
and thus,
\[\bar{V}(k_2)\ge \bar{V}(k_1),\]
as desired. This completes the proof of Step 2.
\end{proof}

\begin{step}
$\bar{V}$ is concave.
\end{step}

\begin{proof}[{\bf Proof of Step 3}]
Fix $k_0,k_1>0$ and $s\in ]0,1[$. Let $(k_i^n(t),c_i^n(t))\in A_{k_i}$ for $i\in \{0,1\}$ such that
\[\int_0^{\infty}e^{-\rho t}u(c_i^n(t),k_i^n(t))dt>\min\{\bar{V}(k_i)-n^{-1},n\}.\]
By Lemma \ref{Lemma5}, we can assume that $c_i^n(t)$ is bounded for $i\in \{0,1\}$. Define
\[c_s^n(t)=(1-s)c_0^n(t)+sc_1^n(t),\]
and consider the following differential equation:
\[\dot{k}(t)=F(k(t),c_s^n(t)),\ k(0)=(1-s)k_0+sk_1\equiv \bar{k}.\]
We show that there exists a solution $k_s^n(t)$ to this equation defined on $\mathbb{R}_+$ such that $k_s^n(t)\ge (1-s)k_0(t)+sk_1(t)$ for every $t\ge 0$.

First, choose any solution $k(t)$ to the above equation defined on an interval $I$ such that $k(t)>0$ for all $t\in I$. Because $F$ is concave, by almost the same proof as in Lemma \ref{Lemma2}, we have that $k(t)\ge (1-s)k_0^n(t)+sk_1^n(t)$ for all $t\in I$. Let $Y$ be the set of all solutions $k(\cdot)$ to the above equation defined on $I_{k(\cdot)}$ such that $k(t)\ge (1-s)k_0^n(t)+sk_1^n(t)$ for all $t\in I_{k(\cdot)}$, where either $I_{k(\cdot)}=[0,T]$ for some $T>0$ or $I_{k(\cdot)}=\mathbb{R}_+$. By the above argument, $Y$ is nonempty. Define a partial order $\succeq$ on $Y$ such that $\hat{k}(\cdot)\succeq \tilde{k}(\cdot)$ if and only if $I_{\tilde{k}(\cdot)}\subset I_{\hat{k}(\cdot)}$ and $\hat{k}(t)=\tilde{k}(t)$ for all $t\in I_{\tilde{k}(\cdot)}$. Let $C$ be a chain in $Y$. If $\cup_{k(\cdot)\in C}I_{k(\cdot)}=\mathbb{R}_+$, then we can define $k^*(t)=k(t)$ for $t\in I_{k(\cdot)}$, and $k^*(\cdot)$ is an upper bound of $C$. If $\sup\cup_{k(\cdot)\in C}I_{k(\cdot)}=T<+\infty$, define $k^*(t)=k(t)$ for $t\in I_{k(\cdot)}$. Then, $k^*(\cdot)$ is a solution to the above equation defined on $[0,T[$. Because $k^*(t)\le k^+(t,\bar{k})$, we can define $k^*(T)=\limsup_{t\to T}k^*(t)\in [0,k^+(T,\bar{k})]$. Define $r(t)$ as in (\ref{RT}), where we use $\max_{t\in [0,T]}k^+(t,\bar{k})$ instead of $\bar{k}$ and $c_s^n(t)$ instead of $c(t)$. Then, $|k^*(t)-k^*(T)|\le \int_t^Tr(\tau)d\tau$, and thus $k^*(\cdot)$ is continuous, and because
\[k^*(t)=\bar{k}+\int_0^tF(k^*(\tau),c_s^n(\tau))d\tau,\]
and $(1-s)k_0^n(t)+sk_1^n(t)$ is continuous, $k^*(\cdot)\in Y$, and thus $k^*(\cdot)$ is an upper bound of $C$. Therefore, by Zorn's lemma, there is a maximal element $k^*(\cdot)\in Y$. Suppose that $I_{k^*(\cdot)}=[0,T]\neq \mathbb{R}_+$. If $k^*(T)>0$, then by Lemma \ref{Lemma2}, there exists $\tilde{k}(\cdot)\in Y$ such that $\tilde{k}(\cdot)\succ k^*(\cdot)$, which contradicts the maximality of $k^*(\cdot)$. Hence, $k^*(T)=k_0^n(T)=k_1^n(T)=0$. Let $c^*=\max\{\sup_{t\ge 0}c_0^n(t),\sup_{t\ge 0}c_1^n(t)\}$, and choose any $k^*>0$ such that $k\mapsto F(k,c)$ is increasing on $[0,k^*]$ for all $c\in [0,c^*]$. Define $r(t)$ as in (\ref{RT}), where we use $k^*$ instead of $\bar{k}$ and $c_s^n(t)$ instead of $c(t)$. Choose a sufficiently small $\varepsilon>0$ such that
\[\max_{i\in \{0,1\},t\in [T,T+\varepsilon]}k_i^n(t)\le k^*,\ \int_T^{T+\varepsilon}r(t)dt\le k^*,\]
and let $I=[T,T+\varepsilon]$ and $X$ be the set of all continuous functions $x:I\to [0,k^*]$ such that $x(T)=0$, $x(t)\ge (1-s)k_0^n(t)+sk_1^n(t)$, and for all $t_1,t_2\in I$ with $t_1<t_2$, $|x(t_2)-x(t_1)|\le \int_{t_1}^{t_2}r(t)dt$. Because $(1-s)k_0^n(\cdot)+sk_1^n(\cdot)\in X$, $X$ is nonempty. The convexity of $X$ is clear. By Ascoli-Arzela's theorem, it is compact with respect to the sup norm. Define
\[P(x(\cdot))(t)=\int_T^tF(x(\tau),c_s^n(\tau))d\tau\]
for all $t\in I$. If $x(\cdot)\in X$ and $y(\cdot)=P(x(\cdot))$, then $y(T)=0$ and
\begin{align*}
y(t)\ge&~\int_T^tF((1-s)k_0^n(\tau)+sk_1^n(\tau),c_s^n(\tau))d\tau\\
\ge&~\int_T^t[(1-s)F(k_0^n(\tau),c_0^n(\tau))+sF(k_1^n(\tau),c_1^n(\tau))]d\tau\\
=&~(1-s)k_0^n(t)+sk_1^n(t)
\end{align*}
and 
\[y(t)\le \int_T^tr(s)ds\le k^*\]
for all $t\in I$. If $T\le t_1<t_2\le T+\varepsilon$, then
\begin{align*}
|y(t_2)-y(t_1)|\le&~\int_{t_1}^{t_2}|F(x(t),c_s^n(t))|dt\\
\le&~\int_{t_1}^{t_2}r(t)dt.
\end{align*}
Therefore, $y(\cdot)\in X$, and thus $P$ is a mapping from $X$ into $X$. Choose any convergent sequence $(x_m(\cdot))$ on $X$ and let $x(\cdot)$ be the limit. Define $y_m(\cdot)=P(x_m(\cdot))$ and $y(\cdot)=P(x(\cdot))$. Fix any $\varepsilon'>0$, and define
\[B_m=\left\{t\in I\left||F(x_m(t),c_s^n(t))-F(x(t),c_s^n(t))|\ge \frac{\varepsilon'}{2\varepsilon}\right.\right\},\]
\[C_m=\cup_{\ell\ge m}B_{\ell}.\]
Because $F$ is continuous, $C_m\downarrow \emptyset$ as $m\to \infty$. Therefore,
\[\lim_{m\to \infty}\int_{C_m}r(t)dt=0.\]
Hence, there exists $m_0\in \mathbb{N}$ such that if $m\ge m_0$, then 
\[\int_{C_m}2r(t)dt<\frac{\varepsilon'}{2}.\]
For every $t\in I$, if $m\ge m_0$, then
\begin{align*}
|y_m(t)-y(t)|\le&~\int_T^{T+\varepsilon}|F(x_m(\tau),c_s^n(\tau))-F(x(\tau),c_s^n(\tau))|d\tau\\
\le&~\int_{I\setminus C_m}|F(x_m(\tau),c_s^n(\tau))-F(x(\tau),c_s^n(\tau))|d\tau\\
&~+\int_{C_m}|F(x_m(\tau),c_s^n(\tau))-F(x(\tau),c_s^n(\tau))|d\tau\\
\le&~\int_{I\setminus C_m}\frac{\varepsilon'}{2\varepsilon}dt+\int_{C_m}2r(\tau)d\tau\\
<&~\frac{\varepsilon'}{2}+\frac{\varepsilon'}{2}=\varepsilon'.
\end{align*}
Hence, $P$ is continuous with respect to the sup norm. By Schauder's fixed point theorem, $P$ has a fixed point $x^*:I\to \mathbb{R}_+$. Define
\[\tilde{k}(t)=\begin{cases}
k^*(t) & \mbox{if }0\le t\le T,\\
x^*(t) & \mbox{if }T\le t\le T+\varepsilon.
\end{cases}\]
Then, $\tilde{k}(\cdot)\in Y$ is an extension of $k^*(\cdot)$, which contradicts the maximality of $k^*(\cdot)$. Hence, the domain of $k^*(\cdot)$ is $\mathbb{R}_+$, and thus we can define $k_s^n(t)$ as $k^*(t)$. 

Thus, for any $M>0$,
\begin{align*}
&~\bar{V}((1-s)k_0+sk_1)\\
\ge&~\int_0^{\infty}e^{-\rho t}u(c_s^n(t),k_s^n(t))dt\\
\ge&~(1-s)\int_0^{\infty}e^{-\rho t}u(c_0^n(t),k_0^n(t))dt+s\int_0^{\infty}e^{-\rho t}u(c_1^n(t),k_1^n(t))dt\\
\ge&\min\{(1-s)\bar{V}(k_0)+s\bar{V}(k_1)-n^{-1},M\},
\end{align*}
for all sufficiently large $n\in \mathbb{N}$, which implies that
\[\bar{V}((1-s)k_0+sk_1)\ge (1-s)\bar{V}(k_0)+s\bar{V}(k_1),\]
as desired. This completes the proof of Step 3.
\end{proof}

Steps 1-3 imply that this proposition is correct. This completes the proof.
\end{proof}
\setcounter{step}{0}

\subsection{The Necessity of the HJB Equation}
First, we show the following result.

\begin{prop}\label{Prop3}
Suppose that Assumptions 1-4 hold. If the value function $\bar{V}$ is finite, then it is increasing, and it is a viscosity solution to the HJB equation. If, in addition, Assumption 5 holds, then $\bar{V}$ is a classical solution to the HJB equation.
\end{prop}

Note that, the examples in Facts 1-3 satisfy Assumptions 1-3 and 5, and in these examples, $\bar{V}$ is finite. Therefore, the violation of Assumption 4 is crucial.

\begin{proof}
We separate the proof into nine steps.

\begin{step}
$\bar{V}$ is increasing.
\end{step}

\begin{proof}[{\bf Proof of Step 1}]
Suppose not. Then, there exist $k_1,k_2>0$ such that $k_1<k_2$ and $\bar{V}(k_1)=\bar{V}(k_2)$. Because $\bar{V}$ is concave and nondecreasing, $\bar{V}(k)=\bar{V}(k_1)$ for every $k>k_1$. 

Choose any $(k(t),c(t))\in A_{k_1}$. Because $u$ is concave, for every $(k^*,c^*)\in \mathbb{R}^2_{++}$,
\begin{align*}
&~\int_0^1e^{-\rho t}[u(c^*,k^*)-u(c(t),k(t))]dt\\
\ge&~\int_0^1e^{-\rho t}\left[\frac{\partial u}{\partial c}(c^*,k^*)(c^*-c(t))+\frac{\partial u}{\partial k}(c^*,k^*)(k^*-k(t))\right]dt\\
=&~\frac{\partial u}{\partial c}(c^*,k^*)\left[\frac{c^*}{\rho}[1-e^{-\rho}]-\int_0^1e^{-\rho t}c(t)dt\right]\\
&~+\frac{\partial u}{\partial k}(c^*,k^*)\left[\frac{k^*}{\rho}[1-e^{-\rho}]-\int_0^1e^{-\rho t}k(t)dt\right]\\
\ge&~\frac{\partial u}{\partial c}(c^*,k^*)\left[\frac{c^*}{\rho}[1-e^{-\rho}]-\int_0^1c(t)dt\right]\\
&~+\frac{\partial u}{\partial k}(c^*,k^*)\left[\frac{k^*}{\rho}[1-e^{-\rho}]-\int_0^1k(t)dt\right].
\end{align*}
Define 
\[N=\max\{F(k,0)|0\le k\le \max_{t\in [0,1]}k^+(t,k_1)\},\]
\[\varepsilon_1=\min\{F(k,0)-F(k,1)|0\le k\le \max_{t\in [0,1]}k^+(t,k_1)\},\]
\[\varepsilon_2=\max\{F(k,0)-F(k,1)|0\le k\le \max_{t\in [0,1]}k^+(t,k_1)\}.\]
Because $F$ is concave, $F(k,1)-F(k,c)\ge \varepsilon_1(c-1)$ for every $k\in [0,\max_{t\in [0,1]}k^+(t,k_1)]$ and $c>1$. Since $0\le k(t)\le k^+(t,k_1)$ for every $t\in [0,1]$,
\begin{align*}
F(k(t),c(t))=&~F(k(t),0)+(F(k(t),c(t))-F(k(t),1))+(F(k(t),1)-F(k(t),0))\\
\le&~N-\varepsilon_1+F(k(t),c(t))-F(k(t),1)\\
\le&~N-\varepsilon_1+\varepsilon_2-\varepsilon_1(c(t)-1)\\
=&~N+\varepsilon_2-\varepsilon_1c(t),
\end{align*}
and thus,
\begin{align*}
0\le&~k(1)\\
=&~k_1+\int_0^1\dot{k}(t)dt\\
=&~k_1+\int_0^1F(k(t),c(t))dt\\
\le&~k_1+\int_0^1(N+\varepsilon_2-\varepsilon_1c(t))dt=k_1+N+\varepsilon_2-\varepsilon_1\int_0^1c(t)dt.
\end{align*}
Therefore,
\[\int_0^1c(t)dt\le \frac{k_1+N+\varepsilon_2}{\varepsilon_1}.\]
Moreover,
\[\int_0^1k(t)dt\le \int_0^1k^+(t,k_1)dt,\]
and thus, for fixed sufficiently large $c^*>0$ and $k^*>k_1$, there exists $\varepsilon>0$ such that
\[\int_0^1e^{-\rho t}[u(c^*,k^*)-u(c(t),k(t))]dt\ge \varepsilon\]
for every $(k(t),c(t))\in A_{k_1}$. Choose any $(k(t),c(t))\in A_{k_1}$ such that $c(t)$ is bounded and
\[\int_0^{\infty}e^{-\rho t}u(c(t),k(t))dt>\bar{V}(k_1)-\frac{\varepsilon}{2},\]
and define
\[\tilde{c}(t)=\begin{cases}
c^* & \mbox{if }0\le t\le 1,\\
c(t) & \mbox{otherwise}.
\end{cases}\]
Choose $\hat{k}>0$ so large that
\[e^{-d_1}\left[\hat{k}-\frac{\delta_2(c^*)}{d_1}(e^{d_1}-1)\right]>k^*.\]
Because $F(k,c^*)\ge -d_1k-\delta_2(c^*)$, by Lemmas \ref{Lemma2} and \ref{Lemma3}, there exists a solution $k(t)$ to the following equation
\[\dot{k}(t)=F(k(t),c^*),\ k(0)=\hat{k}\]
defined on $[0,1]$, and for all $t\in [0,1]$, $k(t)\ge k^*$. By almost the same arguments as in Step 2 of the proof of Proposition \ref{Prop2}, we can show that there exists a function $\tilde{k}(t)$ defined on $\mathbb{R}_+$ such that $\tilde{k}(t)\ge k^*$ for all $t\in [0,1]$, $\tilde{k}(t)\ge k(t)$ for all $t\ge 0$, and $(\tilde{k}(t),\tilde{c}(t))\in A_{\hat{k}}$. Then,
\[\bar{V}(\hat{k})\ge \int_0^{\infty}e^{-\rho t}u(\tilde{c}(t),\tilde{k}(t))dt>\bar{V}(k_1),\]
which is a contradiction. This completes the proof of Step 1.
\end{proof}

\begin{step}
If $\bar{V}$ is differentiable at $\bar{k}>0$, then
\[\sup_{c\ge 0}\{F(\bar{k},c)\bar{V}'(\bar{k})+u(c,\bar{k})\}-\rho \bar{V}(\bar{k})\le 0.\]
\end{step}

\begin{proof}[{\bf Proof of Step 2}]
Fix any $c>0$. By Lemma \ref{Lemma2}, there exists $\varepsilon>0$ such that the differential equation
\[\dot{k}(t)=F(k(t),c),\ k(0)=\bar{k}\]
has a positive and continuously differentiable solution $k(t)$ on $[0,\varepsilon]$. For all $t\in ]0,\varepsilon[$,
\[\int_0^te^{-\rho s}u(c,k(s))ds\le \bar{V}(\bar{k})-e^{-\rho t}\bar{V}(k(t)).\]
Thus,
\[\bar{V}(\bar{k})-\bar{V}(k(t))\ge \int_0^te^{-\rho s}u(c,k(s))ds+(e^{-\rho t}-1)\bar{V}(k(t)).\]
Dividing both sides by $t$ and letting $t\to 0$, we obtain
\[-\bar{V}'(\bar{k})F(\bar{k},c)\ge u(c,\bar{k})-\rho \bar{V}(\bar{k}),\]
which implies that
\[F(\bar{k},c)\bar{V}'(\bar{k})+u(c,\bar{k})-\rho \bar{V}(\bar{k})\le 0.\]
Because the left-hand side is continuous in $c$, the above inequality holds for every $c\ge 0$. Therefore,
\[\sup_{c\ge 0}\{F(\bar{k},c)\bar{V}'(\bar{k})+u(c,\bar{k})\}-\rho \bar{V}(\bar{k})\le 0,\]
as desired. This completes the proof of Step 2.
\end{proof}

Consider the following modified problem of (\ref{MODEL}):
\begin{align}
\max~~~~~&~\int_0^{\infty}e^{-\rho t}u(c(t),k(t))dt \nonumber \\
\mbox{subject to. }&~(k(t),c(t))\in A_{\bar{k}},\label{MODELN}\\
&~c(t)\le n\mbox{ for all }t\ge 0,\nonumber
\end{align}
where $n>0$. Define $A^n_{\bar{k}}=\{(k(t),c(t))\in A_{\bar{k}}|c(t)\le n\mbox{ for all }t\ge 0\}$, and let
\[V_n(\bar{k})=\sup\left\{\left.\int_0^{\infty}e^{-\rho t}u(c(t),k(t))dt\right| (k(t),c(t))\in A^n_{\bar{k}}\right\}.\]
We call the function $V_n$ the value function of the problem (\ref{MODELN}). By almost the same arguments as in the proof of Proposition \ref{Prop2}, we can show that $V_n$ is concave and nondecreasing, and $V_n(\bar{k})>-\infty$ for all $\bar{k}>0$. Because $V_n(\bar{k})\le \bar{V}(\bar{k})<+\infty$, $V_n$ is finite.

\begin{step}
Suppose that $\bar{k}>0$ and $V_n(\bar{k})>M$, and choose any $T>0$ such that
\begin{equation}\label{EV0}
\bar{k}>\frac{\delta_2(n)}{d_1}(e^{d_1T}-1).
\end{equation}
Then, there exists $(\hat{k}(t),\hat{c}(t))\in A^n_{\bar{k}}$ such that $\min_{t\in [0,T]}\hat{k}(t)>0$, $\inf_{t\in [0,T]}\hat{c}(t)>0$, and
\[\int_0^Te^{-\rho t}u(\hat{c}(t),\hat{k}(t))dt>M-e^{-\rho T}V_n(\hat{k}(T)).\]
\end{step}

\begin{proof}[{\bf Proof of Step 3}]
By Assumption 3,
\[F(k,n)\ge -d_1k-\delta_2(n).\]
Therefore, by Lemma \ref{Lemma2}, if $c:[0,T]\to [0,n]$ is measurable, then any solution $k(t)$ to the following differential equation
\begin{equation}\label{KEI}
\dot{k}(t)=F(k(t),c(t))
\end{equation}
satisfies the following inequality:
\begin{equation}\label{EV1}
k(t)\ge e^{-d_1T}\left[\bar{k}-\frac{\delta_2(n)}{d_1}(e^{d_1T}-1)\right]>0.
\end{equation}
By Lemma \ref{Lemma3}, there exists a solution to (\ref{KEI}) defined on $[0,T]$ such that for every $t\in [0,T]$, (\ref{EV1}) holds.

Next, choose any $(k(t),c(t))\in A^n_{\bar{k}}$ such that
\[\int_0^Te^{-\rho t}u(c(t),k(t))dt>M-e^{-\rho T}V_n(k(T)).\]
Define $c_m(t)=\max\{c(t),m^{-1}\}$. As we argued above, there exists a solution $k_m(t)$ to the following ODE:
\[\dot{k}(t)=F(k(t),c_m(t)),\ k(0)=\bar{k}\]
defined on $[0,T]$ that satisfies (\ref{EV1}) for all $t\in [0,T]$. Moreover, because $F$ is Lipschitz on any compact set in $\mathbb{R}_{++}\times \mathbb{R}_+$, there exists $L>0$ such that for every $m\in \mathbb{N}$ and $t\in [0,T]$,
\begin{align*}
|k(t)-k_m(t)|\le&~\int_0^t|F(k(s),c(s))-F(k_m(s),c_m(s))|ds\\
\le&~\int_0^tL[|k(s)-k_m(s)|+|c(s)-c_m(s)|]ds\\
\le&~\int_0^tL[|k(s)-k_m(s)|+m^{-1}]ds.
\end{align*}
Therefore, by Gronwall's inequality,\footnote{See the first part of Lemma \ref{Lemma3} of Hosoya \cite{HOA}.} $\lim_{m\to \infty}\max_{t\in [0,T]}|k(t)-k_m(t)|=0$. We show that
\[\lim_{m\to \infty}\int_0^Te^{-\rho t}u(c_m(t),k_m(t))dt=\int_0^Te^{-\rho t}u(c(t),k(t))dt.\]
Let $h>0$ be sufficiently small. Then, by the monotone convergence theorem,
\[\lim_{m\to \infty}\int_0^Te^{-\rho t}u(c_m(t),k(t)-h)dt=\int_0^Te^{-\rho t}u(c(t),k(t)-h)dt.\]
Hence,
\begin{align*}
&~\liminf_{m\to \infty}\int_0^Te^{-\rho t}u(c_m(t),k_m(t))dt\\
\ge&~\lim_{h\downarrow 0}\int_0^Te^{-\rho t}u(c(t),k(t)-h)dt=\int_0^Te^{-\rho t}u(c(t),k(t))dt.
\end{align*}
By Lemma \ref{Lemma2}, $k_m(t)\le k(t)$ for all $t\in [0,T]$. Thus,
\begin{align*}
&~\limsup_{m\to \infty}\int_0^Te^{-\rho t}u(c_m(t),k_m(t))dt\\
\le&~\limsup_{m\to \infty}\int_0^Te^{-\rho t}u(c_m(t),k(t))dt=\int_0^Te^{-\rho t}u(c(t),k(t))dt,
\end{align*}
which implies that our claim is correct. Hence, if we define $(\hat{k}(t),\hat{c}(t))=(k_m(t),c_m(t))$ on $[0,T]$ for a sufficiently large $m$, choose any $(\tilde{k}(t),\tilde{c}(t))\in A^n_{\hat{k}(T)}$, and define $(\hat{k}(t),\hat{c}(t))=(\tilde{k}(t-T),\tilde{c}(t-T))$ for $t>T$, then all of our claims hold. This completes the proof of Step 3.
\end{proof}

\begin{step}
For every $\bar{k}>0$ and $p\in \partial V_n(\bar{k})$,\footnote{Recall that $p\in \partial V_n(\bar{k})$ if and only if $V_n(k)-V_n(\bar{k})\le p(k-\bar{k})$ for all $k\ge 0$.}
\begin{equation}
\sup_{c\ge 0}\{F(\bar{k},c)p+u(c,\bar{k})\}-\rho V_n(\bar{k})\ge 0. \label{HJBSN}
\end{equation}
\end{step}

\begin{proof}[{\bf Proof of Step 4}]
Choose $T>0$ such that (\ref{EV0}) is satisfied, and fix $t\in ]0,T]$ and $\varepsilon>0$. Because of Step 3, there exists $(k(s),c(s))\in A^n_{\bar{k}}$ such that $\min_{s\in [0,t]}k(s)>0$, $\inf_{s\in [0,t]}c(s)>0$, and
\[\int_0^te^{-\rho s}u(c(s),k(s))ds>V_n(\bar{k})-e^{-\rho t}V_n(k(t))-t\varepsilon.\]
Thus, the mappings $s\mapsto u(c(s),\bar{k})$ and $s\mapsto u(c(s),k(s))$ are integrable on $[0,t]$. By the definition of $p$,
\begin{align*}
\int_0^t-F(k(s),c(s))pds=&~p(\bar{k}-k(t))\\
\le&~V_n(\bar{k})-V_n(k(t))\\
\le&~\int_0^te^{-\rho s}u(c(s),k(s))ds+(e^{-\rho t}-1)V_n(k(t))+t\varepsilon.
\end{align*}
Thus,
\begin{align*}
&~\int_0^t[F(\bar{k},c(s))p+u(c(s),\bar{k})]ds+\int_0^t[u(c(s),k(s))-u(c(s),\bar{k})]ds\\
&~+\int_0^t(e^{-\rho s}-1)u(c(s),k(s))ds+\int_0^t[F(k(s),c(s))-F(\bar{k},c(s))]pds\\
&~+(e^{-\rho t}-1)V_n(k(t))\\
\ge&~-t\varepsilon.
\end{align*}
Because
\[\int_0^t[F(\bar{k},c(s))p+u(c(s),\bar{k})]ds\le \sup_{c\ge 0}\{F(\bar{k},c)p+u(c,\bar{k})\}t,\]
we have
\begin{align*}
&~\sup_{c\ge 0}\{F(\bar{k},c)p+u(c,\bar{k})\}t+(e^{-\rho t}-1)V_n(k(t))\\
\ge&~-t\varepsilon+\int_0^t(1-e^{-\rho s})u(c(s),k(s))ds\\
&~+\int_0^t[F(\bar{k},c(s))-F(k(s),c(s))]pds+\int_0^t[u(c(s),\bar{k})-u(c(s),k(s))]ds\\
\equiv&~-t\varepsilon+J_1(t)+J_2(t)+J_3(t).
\end{align*}
Now,\footnote{Note that, the pair of functions $(k(s),c(s))$ depends on $t>0$. This problem makes the following evaluation difficult.}
\begin{align}
0\ge&~\int_0^t[u(c(s),k(s))-u(n,k^+(s,\bar{k}))]ds\nonumber \\
\ge&~e^{\rho t}\int_0^te^{-\rho s}[u(c(s),k(s))-u(n,k^+(s,\bar{k}))]ds\nonumber \\
\ge&~e^{\rho t}V_n(\bar{k})-V_n(k(t))-e^{\rho t}t\varepsilon-e^{\rho t}\int_0^te^{-\rho s}u(n,k^+(s,\bar{k}))ds\label{EVA}\\
\ge&~e^{\rho t}V_n(\bar{k})-V_n(k^+(t,\bar{k}))-e^{\rho t}t\varepsilon-e^{\rho t}\int_0^te^{-\rho s}u(n,k^+(s,\bar{k}))ds.\nonumber
\end{align}
Moreover,
\begin{align*}
|J_1(t)|\le&~\int_0^t(1-e^{-\rho s})|u(n,k^+(s,\bar{k}))|ds\\
&~+(1-e^{-\rho t})\left|\int_0^t[u(c(s),k(s))-u(n,k^+(s,\bar{k}))]ds\right|,
\end{align*}
and thus, by (\ref{EVA}),
\[\lim_{t\to 0}\frac{J_1(t)}{t}=0.\]
Let $k^*(s)$ be a solution to the following differential equation:
\[\dot{k}(s)=F(k(s),n),\ k(0)=\bar{k}.\]
defined on $[0,t]$. In the proof of Step 3, we have already proved that such a solution exists. Then,
\[|J_2(t)|\le \int_0^tp\max\{|F(k,c)-F(\bar{k},c)||k^*(s)\le k\le k^+(s,\bar{k}),0\le c\le n\}ds,\]
and thus
\[\lim_{t\to 0}\frac{J_2(t)}{t}=0.\]
Finally, because $u$ is concave,
\[u(c(s),\bar{k})-u(c(s),k(s))\ge \frac{\partial u}{\partial k}(c(s),\bar{k})(\bar{k}-k^+(s,\bar{k})),\]
and by Assumption 4, $c\mapsto \frac{\partial u}{\partial k}(c,\bar{k})$ is bounded on $]0,n]$. Therefore,
\[\limsup_{t\to 0}\frac{J_3(t)}{t}\ge 0.\]
In conclusion, we obtain
\[\sup_{c\ge 0}\{F(\bar{k},c)p+u(c,\bar{k})\}-\rho V_n(\bar{k})\ge -\varepsilon,\]
and because $\varepsilon>0$ is arbitrary, (\ref{HJBSN}) holds. This completes the proof of Step 4.
\end{proof}

\begin{step}
For every $\bar{k}>0$, $V_n(\bar{k})\uparrow \bar{V}(\bar{k})$ as $n\to \infty$.
\end{step}

\begin{proof}[{\bf Proof of Step 5}]
This result is immediately obtained from (\ref{FIN}).
\end{proof}

\begin{step}
If $\bar{V}$ is differentiable at $\bar{k}>0$, then
\[\sup_{c\ge 0}\{F(\bar{k},c)\bar{V}'(\bar{k})+u(c,\bar{k})\}-\rho \bar{V}(\bar{k})\ge 0.\]
\end{step}

\begin{proof}[{\bf Proof of Step 6}]
For any sufficiently large $n$, choose any $p_n\in \partial V_n(\bar{k})$. Fix any $\delta>0$, and choose $k_1,k_2>0$ with $k_1<\bar{k}<k_2$ and
\[\bar{V}'(\bar{k})-\delta<\frac{\bar{V}(k_2)-\bar{V}(\bar{k})}{k_2-\bar{k}}\le \bar{V}'(\bar{k})\le \frac{\bar{V}(k_1)-\bar{V}(\bar{k})}{k_1-\bar{k}}<\bar{V}'(\bar{k})+\delta.\]
Because $V_n(k_i)\uparrow \bar{V}(k_i)$ as $n\to \infty$ for $i=1,2$ and $V_n(\bar{k})\uparrow \bar{V}(\bar{k})$,
\[\max_{i=1,2}\left|\frac{V_n(k_i)-V_n(\bar{k})}{k_i-\bar{k}}-\frac{\bar{V}(k_i)-\bar{V}(\bar{k})}{k_i-\bar{k}}\right|<\delta\]
for any sufficiently large $n$. Therefore,
\[|p_n-\bar{V}'(\bar{k})|\le 2\delta\]
for sufficiently large $n$, and thus $\lim_{n\to \infty}p_n=\bar{V}'(\bar{k})$. 

Because $\bar{V}$ is increasing, $\bar{V}'(\bar{k})>0$. Hence, $p_n>0$ for sufficiently large $n$ and
\begin{align*}
0\le &~\sup_{c\ge 0}\{F(\bar{k},c)p_n+u(c,\bar{k})\}-\rho V_n(\bar{k})\\
=&~F(\bar{k},c^*(p_n,\bar{k}))p_n+u(c^*(p_n,\bar{k}),\bar{k})-\rho V_n(\bar{k})\\
\to&~F(\bar{k},c^*(\bar{V}'(\bar{k}),\bar{k}))\bar{V}'(\bar{k})+u(c^*(\bar{V}'(\bar{k}),\bar{k}),\bar{k})-\rho \bar{V}(\bar{k})\\
=&~\sup_{c\ge 0}\{F(\bar{k},c)\bar{V}'(\bar{k})+u(c,\bar{k})\}-\rho \bar{V}(\bar{k}),
\end{align*}
as desired. This completes the proof of Step 6.
\end{proof}

\begin{step}
For all $\bar{k}>0$ and $p\in \partial \bar{V}(\bar{k})$,
\[\sup_{c\ge 0}\{F(\bar{k},c)p+u(c,\bar{k})\}-\rho \bar{V}(\bar{k})=0.\]
\end{step}

\begin{proof}[{\bf Proof of Step 7}]
First, let $p_1=D_-\bar{V}(k), k_m\uparrow k$ as $m\to \infty$, and $\bar{V}'(k_m)$ can be defined for all $m$. Because $\bar{V}$ is concave, $\bar{V}'(k_m)$ is nonincreasing in $m$, and thus converges to some number $p^*\ge p_1$ as $m\to \infty$. If $p^*>p_1$, then there exists $k'<k$ such that
\[\frac{\bar{V}(k')-\bar{V}(k)}{k'-k}<p^*.\]
Because $k_m\uparrow k$ as $m\to \infty$, there exists $m$ such that
\[k_m>k',\ \frac{\bar{V}(k')-\bar{V}(k_m)}{k'-k_m}<p^*.\]
This implies that $\bar{V}'(k_m)<p^*$, which is a contradiction. Therefore, $\bar{V}'(k_m)$ converges to $p_1$. Similarly, if $p_2=D_+\bar{V}(k), k_m\downarrow k$ as $m\to \infty$ and $\bar{V}'(k_m)$ can be defined for all $m$, then $\bar{V}'(k_m)$ converges to $p_2$ as $m\to \infty$. In particular, if $\bar{V}$ is differentiable around $\bar{k}$, then $\bar{V}'$ is continuous at $\bar{k}$.

Choose any $\bar{k}>0$. If $\bar{V}$ is differentiable at $\bar{k}$, then $\partial \bar{V}(\bar{k})=\{\bar{V}'(\bar{k})\}$, and the claim of Step 7 follows from Steps 2 and 6. Therefore, suppose that $\bar{V}$ is not differentiable at $\bar{k}$, and $p_1=D_-\bar{V}(\bar{k}), p_2=D_+\bar{V}(\bar{k})$. By assumption and Step 1, $p_1>p_2>0$ and $\partial \bar{V}(\bar{k})=[p_2,p_1]$. Choose sequences $(k_1^m), (k_2^m)$ such that $k_1^m\uparrow \bar{k}, k_2^m\downarrow \bar{k}$ as $m\to \infty$, and $\bar{V}$ is differentiable at $k_i^m$ for all $m$. By the above arguments, $\bar{V}'(k_1^m)\downarrow p_1$ and $\bar{V}'(k_2^m)\uparrow p_2$ as $m\to \infty$. By Steps 2 and 6,
\[\sup_{c\ge 0}\{F(\bar{k},c)p_i+u(c,\bar{k})\}-\rho \bar{V}(\bar{k})=0.\]
Next, choose any $p$ with $p_2<p<p_1$, and define $\varepsilon=\min\{p_1-p,p-p_2\}$. As in the proof of Step 6, we can show that for any $m$ and $i\in \{1,2\}$,
\[q\in \partial V_n(k_i^m)\Rightarrow |q-\bar{V}'(k_i^m)|<\varepsilon\]
for sufficiently large $n$. Choose $n_m$ as such an $n$. We can assume that $n_m$ is increasing in $m$. Because $|\bar{V}'(k_i^m)-p|\ge \varepsilon$, there exists $k^m\in [k_1^m,k_2^m]$ such that $p\in \partial V_{n_m}(k^m)$. Then, by Step 4,
\[\sup_{c\ge 0}\{F(k^m,c)p+u(c,k^m)\}-\rho V_{n_m}(k^m)\ge 0.\]
Because $V_{n_m}(k)\uparrow \bar{V}(k)$ as $m\to \infty$ pointwise, by Dini's theorem, $V_{n_m}\to \bar{V}$ uniformly as $m\to \infty$ on any compact set in $\mathbb{R}_{++}$. Therefore,
\[\sup_{c\ge 0}\{F(\bar{k},c)p+u(c,\bar{k})\}-\rho \bar{V}(\bar{k})\ge 0.\]
Now, define
\[g(p)=\sup_{c\ge 0}\{F(\bar{k},c)p+u(c,\bar{k})\}.\]
Suppose that $p,q\in [p_2,p_1]$ and $r=(1-t)p+tq$ for $t\in [0,1]$. Then,
\begin{align*}
g(r)=&~F(\bar{k},c^*(r,\bar{k}))r+u(c^*(r,\bar{k}),\bar{k})\\
=&~(1-t)[F(\bar{k},c^*(r,\bar{k}))p+u(c^*(r,\bar{k}),\bar{k})]+t[F(\bar{k},c^*(r,\bar{k}))q+u(c^*(r,\bar{k}),\bar{k})]\\
\le &~(1-t)g(p)+tg(q).
\end{align*}
Therefore, $g$ is convex. Suppose that $p\in [p_2,p_1]$. Because $g(p_1)=g(p_2)=\rho \bar{V}(\bar{k})$ and $g(p)\ge \rho \bar{V}(\bar{k})$, we have that $g(p)=\rho \bar{V}(\bar{k})$. This completes the proof of Step 7.
\end{proof}

\begin{step}
$\bar{V}$ is a viscosity solution to the HJB equation.
\end{step}

\begin{proof}[{\bf Proof of Step 8}]
Choose any $\bar{k}>0$. Suppose that $\varphi(k)$ is a continuously differentiable function defined on a neighborhood of $\bar{k}$ such that $\varphi(\bar{k})=\bar{V}(\bar{k})$ and either $\varphi(k)\ge \bar{V}(k)$ for all $k$ or $\varphi(k)\le \bar{V}(k)$ for all $k$. Then, $\bar{k}$ is an extremal point of $\varphi(k)-\bar{V}(k)$, and thus $\varphi'(\bar{k})\in \partial \bar{V}(\bar{k})$. This implies that
\[\sup_{c\ge 0}\{F(\bar{k},c)\varphi'(\bar{k})+u(c,\bar{k})\}-\rho \bar{V}(\bar{k})=0.\]
Therefore, $\bar{V}$ is a viscosity solution to the HJB equation. This completes the proof of Step 8.
\end{proof}

\begin{step}
If Assumption 5 holds, then $\bar{V}$ is continuously differentiable.
\end{step}

\begin{proof}[{\bf Proof of Step 9}]
As we argued in the proof of Step 7, it suffices to show that $\bar{V}$ is differentiable at any $\bar{k}>0$. Suppose that $\bar{V}$ is not differentiable at $\bar{k}>0$, and let $p_1=D_-\bar{V}(\bar{k})$ and $p_2=D_+\bar{V}(\bar{k})$. Then, $p_2<p_1$. Define a function $g$ as in the proof of Step 7, and choose $p\in ]p_2,p_1[$. Then, for every $q$ around $p$,
\[g(q)\ge F(\bar{k},c^*(p,\bar{k}))q+u(c^*(p,\bar{k}),\bar{k}),\ g(p)=F(\bar{k},c^*(p,\bar{k}))p+u(c^*(p,\bar{k}),\bar{k}).\]
Because $g$ is constant around $p$, we must have $F(\bar{k},c^*(p,\bar{k}))=0$ for such $p$. Therefore, there exists $c^*>0$ such that $c^*(p,\bar{k})=c^*$ for every $p\in ]p_2,p_1[$. However, this implies that $\frac{\partial F}{\partial c}(\bar{k},c^*)p+\frac{\partial u}{\partial c}(c^*,\bar{k})=0$ for every $p\in ]p_2,p_1[$, which is impossible. This completes the proof of Step 9.
\end{proof}

Steps 1-9 show that all the claims in Proposition \ref{Prop3} are correct. This completes the proof.
\end{proof}
\setcounter{step}{0}

Because Proposition \ref{Prop3} requires the finiteness of the value function, we want an additional assumption that ensures the finiteness of the value function. First, define the CRRA function as follows:
\[u_{\theta}(x)=\begin{cases}
\frac{x^{1-\theta}-1}{1-\theta} & \mbox{if }\theta\neq 1,\\
\log x & \mbox{if }\theta=1,
\end{cases}\]
where $\theta>0$.\footnote{This function is the solution to the following differential equation:
\[-x\frac{u''(x)}{u'(x)}=\theta,\ u(1)=0,\ u'(1)=1,\]
where the left-hand side is sometimes called the relative risk aversion of $u$. Therefore, this function is called the \textbf{constant relative risk aversion} function, and abbreviatedly, the CRRA function.}

\vspace{12pt}
\noindent
\textbf{Assumption 6}. There exist $k^*>0, c^*\ge 0, \gamma>0, \delta>0, \theta>0, a>0, b\ge 0, C\in \mathbb{R}$ such that
\begin{align}
&~(\gamma,-\delta)\in \partial F(k^*,c^*),\label{FL}\\
&~\rho-(1-\theta)\gamma>0,\label{SL}\\
&~u(c,k)\le au_{\theta}(c)+bu_{\theta}(k)+C\mbox{ for all }c>0, k>0.\label{TL}
\end{align}

\vspace{12pt}
We stress that Assumption 6 is not strong. For example, in the RCK model, $F(k,c)=f(k)-c$ and $u(c,k)=u(c)$. Suppose that $f(k)$ is differentiable. Then, the condition $(\gamma,-\delta)\in \partial F(k^*,c^*)$ means that $\gamma=f'(k^*)$ and $\delta=1$. Hence, requirement (\ref{FL}) is satisfied if $k^*$ is sufficiently small and $c^*=0$. Therefore, the actual restrictions are only (\ref{SL}) and (\ref{TL}). Moreover, if $f$ satisfies the Inada condition, we can choose $\gamma$ so small that $\rho-(1-\theta)\gamma>0$, and thus the actual restriction is the existence of $\theta>0$ such that $u(c)\le au_{\theta}(c)+C$ for all $c>0$. This is just a mild restriction on $u$. Additionally, if we can choose $\theta\ge 1$, then (\ref{SL}) is automatically satisfied and vanishes even when $f(k)$ does not satisfy the Inada condition.

Note that, if $u(c,k)=u_{\theta}(c)$ and $F(k,c)=\gamma k-c$, then (\ref{SL}) provides a necessary and sufficient condition for the finiteness of the value function.\footnote{The technology described by the function $F(k,c)=\gamma k-c$ is called \textbf{AK technology}.} Sufficiency is proved in Ch.4 of Barro and Sara-i-Martin \cite{BS1}, and necessity is shown in Hosoya and Kuwata \cite{HK}. Therefore, we think that (\ref{SL}) is a crucial requirement for the finiteness of the value function.

On the other hand, if $\sup_{c,k}u(c,k)<+\infty$, the requirement (\ref{TL}) is automatically satisfied for every $\theta\in ]0,1[$, because $u_{\theta}(0)>-\infty$ when $0<\theta<1$. Therefore, again our requirements are automatically satisfied. Next, suppose that $u(c,k)=(c^{\sigma}+k^{\sigma})^{\frac{A}{\sigma}}$ and $\sigma<1,\ \sigma\neq 0,\ 0<A<1$. If $\sigma>0$, then
\[u(c,k)\le (c^{\sigma}+k^{\sigma})^{\frac{A-1}{\sigma}}2^{\frac{1}{\sigma}}(c+k)\le 2^{\frac{1}{\sigma}}(c^A+k^A),\]
and thus it satisfies (\ref{TL}) for $\theta=1-A$. If $\sigma<0$, then because $x\mapsto x^{\frac{A}{\sigma}}$ is decreasing, by the formula for arithmetic-geometric means,
\[u(c,k)\le 2^{\frac{A}{\sigma}}(ck)^{\frac{A}{2}}\le 2^{\frac{A}{\sigma}-1}(c^A+k^A),\]
and thus it satisfies (\ref{TL}) for $\theta=1-A$. Therefore, there are sufficiently many models that satisfy Assumption 6.

The following is our first main result.\footnote{Recall that $\mathscr{V}$ is the space of all functions $V:\mathbb{R}_{++}\to \mathbb{R}$ that is increasing and concave, and satisfies the growth condition.}

\begin{thm}\label{Theorem1}
Suppose that Assumptions 1-6 hold. Then $\bar{V}\in \mathscr{V}$, and $\bar{V}$ is a classical solution to the HJB equation.
\end{thm}

\begin{proof}
First, consider the following problem:
\begin{align*}
\max~~~~~&~\int_0^{\infty}e^{-\rho t}u(c(t),k(t))dt\\
\mbox{subject to. }&~c(t)\in W_1,\\
&~k(t)\ge 0,\ c(t)\ge 0,\\
&~\int_0^{\infty}e^{-\rho t}u(c(t),k(t))dt\mbox{ can be defined},\\
&~\dot{k}(t)=\gamma (k(t)-k^*)-\delta (c(t)-c^*)+F(k^*,c^*)\mbox{ a.e.},\\
&~k(0)=\bar{k}.
\end{align*}
Define $A^L_{\bar{k}}$ as the set of all pairs $(k(t),c(t))$ of nonnegative functions such that $k(t)$ is absolutely continuous on every compact set, $c(t)\in W_1$, $\int_0^{\infty}e^{-\rho t}u(c(t),k(t))dt$ can be defined, $k(0)=\bar{k}$ and
\[\dot{k}(t)=\gamma (k(t)-k^*)-\delta(c(t)-c^*)+F(k^*,c^*)\]
for almost all $t\ge 0$. Let
\[V_1(\bar{k})=\sup\left\{\left.\int_0^{\infty}e^{-\rho t}u(c(t),k(t))dt\right|(k(t),c(t))\in A^L_{\bar{k}}\right\}.\]

\begin{step}
$V_1(\bar{k})\ge \bar{V}(\bar{k})$ for all $\bar{k}>0$.
\end{step}

\begin{proof}[{\bf Proof of Step 1}]
By Proposition \ref{Prop2}, $\bar{V}(\bar{k})>-\infty$. For every $\varepsilon>0$ and $N>0$, there exists $(k(t),c(t))\in A_{\bar{k}}$ such that $c(t)$ is bounded and
\[\int_0^{\infty}e^{-\rho t}u(c(t),k(t))dt\ge \min\{\bar{V}(\bar{k})-\varepsilon,N\}.\]
Consider the following differential equation:
\[\dot{k}(t)=\gamma (k(t)-k^*)-\delta (c(t)-c^*)+F(k^*,c^*),\ k(0)=\bar{k}.\]
The solution to the above equation is:
\[\hat{k}(t)=e^{\gamma t}\left[\bar{k}-\int_0^te^{-\gamma s}(\gamma k^*+\delta(c(s)-c^*)-F(k^*,c^*))ds\right].\]
Because $(\gamma,-\delta)\in \partial F(k^*,c^*)$, $F(k,c)\le \gamma (k-k^*)-\delta (c-c^*)+F(k^*,c^*)$ for all $(k,c)$, and thus, by Lemma \ref{Lemma2}, $\hat{k}(t)\ge k(t)$ for every $t\ge 0$. Therefore, $(\hat{k}(t),c(t))\in A^L_{\bar{k}}$, and thus
\[V_1(\bar{k})\ge \min\{\bar{V}(\bar{k})-\varepsilon,N\}.\]
Because $\varepsilon, N$ are arbitrary, we have that $V_1(\bar{k})\ge \bar{V}(\bar{k})$. This completes the proof of Step 1.
\end{proof}

Second, consider the following problem: 
\begin{align*}
\max~~~~~&~\int_0^{\infty}e^{-\rho t}[au_{\theta}(c(t))+bu_{\theta}(k(t))]dt\\
\mbox{subject to. }&~c(t)\in W_1,\\
&~k(t)\ge 0,\ c(t)\ge 0,\\
&~\int_0^{\infty}e^{-\rho t}[au_{\theta}(c(t))+bu_{\theta}(k(t))]dt\mbox{ can be defined},\\
&~\dot{k}(t)=\gamma (k(t)-k^*)-\delta(c(t)-c^*)+F(k^*,c^*)\mbox{ a.e.},\\
&~k(0)=\bar{k}.
\end{align*}
Define $A^{L2}_{\bar{k}}$ as the set of all pairs $(k(t),c(t))$ of nonnegative functions such that $k(t)$ is absolutely continuous in every compact set, $c(t)\in W_1$, $\int_0^{\infty}e^{-\rho t}[au_{\theta}(c(t))+bu_{\theta}(k(t))]dt$ can be defined, $k(0)=\bar{k}$ and
\[\dot{k}(t)=\gamma(k(t)-k^*)-\delta(c(t)-c^*)+F(k^*,c^*)\]
for almost all $t\ge 0$. Let
\[V_2(\bar{k})=\sup\left\{\left.\int_0^{\infty}e^{-\rho t}[au_{\theta}(c(t))+bu_{\theta}(k(t))]dt\right|(k(t),c(t))\in A^{L2}_{\bar{k}}\right\}.\]

\begin{step}
$V_2(\bar{k})+\frac{C}{\rho}\ge V_1(\bar{k})$ for all $\bar{k}>0$.
\end{step}

\begin{proof}[{\bf Proof of Step 2}]
By Proposition \ref{Prop2} and Step 1, $V_1(\bar{k})>-\infty$. For every $\varepsilon>0$ and $N>0$, there exists $(k(t),c(t))\in A^L_{\bar{k}}$ such that
\[\int_0^{\infty}e^{-\rho t}u(c(t),k(t))dt\ge \min\{V_1(\bar{k})-\varepsilon,N\}.\]
Because
\[\int_0^{\infty}e^{-\rho t}\min\{au_{\theta}(c(t))+bu_{\theta}(k(t))+C,0\}dt\ge \int_0^{\infty}e^{-\rho t}\min\{u(c(t),k(t)),0\}dt>-\infty,\]
we have that
\[\int_0^{\infty}e^{-\rho t}[au_{\theta}(c(t))+bu_{\theta}(k(t))]dt\]
can be defined. Hence, $(k(t),c(t))\in A^{L2}_{\bar{k}}$, and thus,
\[V_2(\bar{k})+\frac{C}{\rho}\ge \min\{V_1(\bar{k})-\varepsilon,N\}.\]
Because $\varepsilon, N$ are arbitrary, $V_2(\bar{k})+\frac{C}{\rho}\ge V_1(\bar{k})$. This completes the proof of Step 2.
\end{proof}

Define
\[\hat{k}=\bar{k}-k^*+\frac{\delta c^*+F(k^*,c^*)}{\gamma},\ C^*=\frac{\rho-(1-\theta)\gamma}{\theta \delta}\hat{k}.\]
Note that,
\[-k^*+\frac{\delta c^*+F(k^*,c^*)}{\gamma}=\frac{1}{\gamma}[\gamma(0-k^*)-\delta(0-c^*)+F(k^*,c^*)]\ge 0,\]
and thus $\hat{k},C^*>0$ for all $\bar{k}>0$. Define
\[V_3(\bar{k})=\begin{cases}
\frac{(C^*)^{1-\theta}\theta}{(1-\theta)(\rho-(1-\theta)\gamma)}-\frac{1}{\rho(1-\theta)}, & \mbox{if }\theta\neq 1,\\
\frac{\log C^*}{\rho}+\frac{\gamma-\rho}{\rho^2} & \mbox{if }\theta=1
\end{cases}\]
and
\[V_4(\bar{k})=\begin{cases}
\frac{\hat{k}^{1-\theta}}{(1-\theta)(\rho-(1-\theta)\gamma)}-\frac{1}{\rho(1-\theta)} & \mbox{if }\theta\neq 1,\\
\frac{\log \hat{k}}{\rho}+\frac{\gamma}{\rho^2}, & \mbox{if }\theta=1.
\end{cases}\]

\begin{step}
$aV_3(\bar{k})+bV_4(\bar{k})\ge V_2(\bar{k})$ for all $\bar{k}>0$.
\end{step}

\begin{proof}[{\bf Proof of Step 3}]
Let
\[c^*(t)=C^*e^{\frac{\gamma-\rho}{\theta}t},\]
\[k^*(t)=e^{\gamma t}\left[\bar{k}-\int_0^te^{-\gamma s}[\gamma k^*+\delta(c^*(s)-c^*)-F(k^*,c^*)]ds\right].\]
Note that,
\[\dot{k}^*(t)=\gamma (k^*(t)-k^*)-\delta (c^*(t)-c^*)+F(k^*,c^*),\]
\[\frac{d}{dt}(u_{\theta}'(c^*(t)))=(\rho-\gamma)u_{\theta}'(c^*(t)).\]
Then, for every $(k(t),c(t))\in A^{L2}_{\bar{k}}$,
\begin{align*}
&~\int_0^Te^{-\rho t}(u_{\theta}(c(t))-u_{\theta}(c^*(t)))dt\\
\le&~\int_0^Te^{-\rho t}u_{\theta}'(c^*(t))(c(t)-c^*(t))dt\\
=&~\delta^{-1}\int_0^Te^{-\rho t}u_{\theta}'(c^*(t))[\gamma (k(t)-k^*(t))-(\dot{k}(t)-\dot{k}^*(t))]dt\\
=&~\delta^{-1}e^{-\rho T}u_{\theta}'(c^*(T))(k^*(T)-k(T))\\
\le&~\delta^{-1}e^{-\rho T}u_{\theta}'(c^*(T))k^*(T)\\
=&~\delta^{-1}(C^*)^{-\theta}\left[\bar{k}-\left(k^*-\frac{\delta c^*+F(k^*,c^*)}{\gamma}\right)(1-e^{-\gamma T})-\frac{\theta \delta C^*}{\rho-(1-\theta)\gamma}(1-e^{\frac{(1-\theta)\gamma-\rho}{\theta}T})\right]\\
\to&~0\mbox{ (as $T\to \infty$)}.
\end{align*}
Moreover,
\begin{align*}
k(t)=&~e^{\gamma t}\left(\bar{k}-\int_0^te^{-\gamma s}[\gamma k^*+\delta(c(s)-c^*)-F(k^*,c^*)]ds\right)\\
\le&~e^{\gamma t}\left(\bar{k}-\int_0^te^{-\gamma s}[\gamma k^*-\delta c^*-F(k^*,c^*)]ds\right)\\
\le&~e^{\gamma t}\left(\bar{k}-\int_0^{\infty}e^{-\gamma s}[\gamma k^*-\delta c^*-F(k^*,c^*)]ds\right)\\
=&~\left(\bar{k}-k^*+\frac{\delta c^*+F(k^*,c^*)}{\gamma}\right)e^{\gamma t}=\hat{k}e^{\gamma t},
\end{align*}
and thus,
\begin{align*}
&~\int_0^{\infty}e^{-\rho t}[au_{\theta}(c(t))+bu_{\theta}(k(t))]dt\\
\le&~a\int_0^{\infty}e^{-\rho t}u_{\theta}(c^*(t))dt+b\int_0^{\infty}e^{-\rho t}u_{\theta}(\hat{k}e^{\gamma t})dt\\
=&~aV_3(\bar{k})+bV_4(\bar{k}),
\end{align*}
which completes the proof.
\end{proof}

By Steps 1-3, $\bar{V}(\bar{k})<+\infty$ for every $\bar{k}>0$.

\begin{step}
For every $\bar{k}>0$, $e^{-\rho t}\bar{V}(k^+(t,\bar{k}))\to 0$ as $t\to \infty$.
\end{step}

\begin{proof}[{\bf Proof of Step 4}]
We have already shown that $\inf_{t\ge 0}k^+(t,\bar{k})>0$ for all $\bar{k}>0$, and thus it suffices to show that for $i\in \{3,4\}$,
\[\limsup_{t\to\infty}e^{-\rho t}V_i(k^+(t,\bar{k}))\le 0,\]
Define
\[\hat{k}(t)=\hat{k}e^{\gamma t}.\]
By Lemma \ref{Lemma2} and the calculation in Step 3, $\hat{k}(t)\ge k^+(t,\bar{k})$ for all $t\ge 0$, and it suffices to show that for $i\in \{3,4\}$,
\[\lim_{t\to \infty}e^{-\rho t}V_i(\hat{k}(t))=0.\]
If $\theta=1$, then 
\[V_i(\hat{k}(t))=A\log (e^{\gamma t}+B)+C\]
for some $A>0,\ B\ge 0$, and $C\in\mathbb{R}$, and
\[e^{-\rho t}V_i(\hat{k}(t))=Ae^{-\rho t}\log (e^{\gamma t}+B)+e^{-\rho t}C\to 0\]
as $t\to \infty$. If $\theta\neq 1$, then
\[V_i(\hat{k}(t))=A(e^{\gamma t}+B)^{1-\theta}+C,\]
for some $A, B, C\in\mathbb{R}$ such that $A(1-\theta)>0$ and $B\ge 0$. If $\theta<1$, then
\[e^{-\rho t}V_i(\hat{k}(t))=A(e^{\frac{(1-\theta)\gamma-\rho}{1-\theta}t}+e^{-\frac{\rho}{1-\theta}t}B)^{1-\theta}+e^{-\rho t}C\to 0\]
as $t\to\infty$. If $\theta>1$, then
\[e^{-\rho t}V_i(\hat{k}(t))=e^{-\rho t}[A(e^{\gamma t}+B)^{1-\theta}+C]\to 0\]
as $t\to \infty$. Thus, in any case, our claim is correct. This completes the proof of Step 4.
\end{proof}

Step 4 states that all of our claims in Theorem \ref{Theorem1} are correct. This completes the proof.
\end{proof}
\setcounter{step}{0}

We make some remarks on Theorem \ref{Theorem1}. First, in many related studies on the HJB equation in control problems, $u$ is assumed to be bounded. Under this condition, the value function is automatically bounded and finite, and thus there are many techniques that can be used in the proof. See, for example, Ch.3 of Bardi and Capuzzo-Dolcetta \cite{BCD}. However, we want to consider the case in which $u(c)=\log c$, and the logarithmic function is not bounded. Hence, we cannot assume the boundedness of $u$ in this paper. Note that $u(c)=\log c$ is not excluded from our assumptions: set $\theta=1$ and check (\ref{SL}) and (\ref{TL}) of Assumption 6.

In Theorem \ref{Theorem1}, the value function automatically satisfies the growth condition (\ref{GC}). If (\ref{MODEL}) violates Assumptions 1-6, then (\ref{GC}) may be not satisfied even when the value function is a classical solution to the HJB equation. For example, let $\rho=1, u(c,k)=c$, and $F(k,c)=k-c$. Then, we have already obtained the value function $\bar{V}(\bar{k})=\bar{k}$ in the proof of Fact 1. Because $k^+(t,\bar{k})=\bar{k}e^t$, (\ref{GC}) is violated. However, we think that this is a pathological example.

Meanwhile, if $\rho=2, u(c,k)=c$ and $F(k,c)=k-c$, then we have already shown in Fact 1 that $0\le \bar{V}(\bar{k})\le \bar{k}$ and $k^+(t,\bar{k})=\bar{k}e^t$, which implies that $\bar{V}$ satisfies (\ref{GC}). However, in this case $\bar{V}$ does not solve the HJB equation. Therefore, we have also obtained a model in which $\bar{V}\in\mathscr{V}$ and $\bar{V}$ violates the HJB equation. This is another pathological example.

\subsection{Sufficiency of the HJB Equation and Construction of the Solution}

In this subsection, we examine the sufficiency of the HJB equation for a function to be the value function. We need an additional assumption.

\vspace{12pt}
\noindent
\textbf{Assumption 7}. Both $\frac{\partial F}{\partial c}$ and $\frac{\partial u}{\partial c}$ are continuously differentiable in $k$ on $\mathbb{R}^2_{++}$. Moreover, there exists $k>0$ such that $\inf_{c\ge 0}D_{k,+}F(k,c)>\rho$.

\vspace{12pt}
Note that, in the RCK model, $u(c,k)=u(c)$ and $F(k,c)=f(k)-c$. Thus, the first assertion of Assumption 7 is automatically satisfied even if $f(k)$ is not differentiable. If, additionally, there exists $k$ such that $f'(k)>\rho$, then the second assertion of Assumption 7 is also satisfied. Therefore, this assumption is also not strong.

We define additional notation. Let $B_{\bar{k}}$ denote the set of all pairs of nonnegative functions $(k(t),c(t))$ defined on $\mathbb{R}_+$ such that $k(t)$ is absolutely continuous on any compact interval, $c(t)\in W$, $\lim_{T\to \infty}\int_0^Te^{-\rho t}u(c(t),k(t))dt$ exists, $k(0)=\bar{k}$, and
\[\dot{k}(t)=F(k(t),c(t))\]
for almost all $t\ge 0$. Clearly $A_{\bar{k}}\subset B_{\bar{k}}$, but it is unknown whether $A_{\bar{k}}=B_{\bar{k}}$.

The next proposition is crucial for our next main result.

\begin{prop}\label{Prop4}
Suppose that Assumptions 1-5 and 7 hold. Suppose also that $V\in \mathscr{V}$ is a classical solution to the HJB equation. Choose any $\bar{k}>0$, and consider the following differential equation:\footnote{For the definition of the function $c^*(p,k)$, see Proposition \ref{Prop1}.}
\begin{equation}
\dot{k}(t)=F(k(t),c^*(V'(k(t)),k(t))),\ k(0)=\bar{k}.\label{SOL}
\end{equation}
Then, there exists a solution $k^*(t)$ to the above equation defined on $\mathbb{R}_+$, and for any such solution, $\inf_{t\ge 0}k^*(t)>0$. If we define $c^*(t)=c^*(V'(k^*(t)),k^*(t))$, then $c^*(t)$ is continuous and $(k^*(t),c^*(t))\in B_{\bar{k}}$. Moreover, 
\[V(\bar{k})=\lim_{T\to \infty}\int_0^Te^{-\rho t}u(c^*(t),k^*(t))dt,\]
and for every $(k(t),c(t))\in B_{\bar{k}}$, if $\inf_{t\ge 0}k(t)>0$, then 
\[\lim_{T\to \infty}\int_0^Te^{-\rho t}u(c(t),k(t))dt\le V(\bar{k}).\]
\end{prop}

\begin{proof}

We separate the proof into three steps.

\begin{step}
Suppose that $V\in\mathscr{V}$ is a classical solution to the HJB equation. Then, there exists a solution $k^*(t)$ to the differential equation $(\ref{SOL})$ defined on $\mathbb{R}_+$. Moreover, for any such solution, $\inf_{t\ge 0}k^*(t)>0$.
\end{step}

\begin{proof}[{\bf Proof of Step 1}]
First, note that (\ref{SOL}) is an autonomous ODE such that the right-hand side is continuous with respect to $k$.\footnote{That is, if we define $g(k)=F(k,c^*(\bar{V}'(k),k))$, then (\ref{SOL}) is equivalent to
\[\dot{k}(t)=g(k(t)),\ k(0)=\bar{k}.\]
Because the function $g(k)$ is continuous in $k$ and independent of $t$, all requirements in i) of Lemma \ref{Lemma2} are satisfied.} Applying Lemma \ref{Lemma2} for (\ref{SOL}), there exists $T>0$ and a solution $k(t)$ to (\ref{SOL}) defined on $[0,T]$ that is continuously differentiable.

Let $\varepsilon>0$ satisfy that
\[\inf_{c\ge 0}D_{k,+}F(\varepsilon,c)>\rho,\ \varepsilon<\bar{k}.\]
We show that for any solution $k(t)$ to (\ref{SOL}) defined on $[0,T]$, $k(t)\ge \varepsilon$.

Suppose that this claim is incorrect. Then, there exists $t^*>0$ such that $0<k(t^*)<\varepsilon$. By the mean value theorem, we can assume that $t^*<T$ and $\dot{k}(t^*)<0$. By Alexandrov's theorem, $V$ is twice differentiable almost everywhere, and thus we can assume that $V''(k(t^*))$ can be defined.\footnote{See Alexandrov \cite{AL} or Howard \cite{HO}.} Define
\[c(t)=c^*(V'(k(t)),k(t))\]
on $[0,T]$. Then, $c(t)$ is continuous and positive. Because $V$ is a classical solution to the HJB equation,
\[F(k(t),c(t))V'(k(t))+u(c(t),k(t))=\rho V(k(t)),\]
for every $t$ around $t^*$. Because $\dot{k}(t^*)<0$, for any sufficiently small $h>0$, there exists $k_1,k_2\in [k(t^*+h),k(t^*)],\ \theta_1,\theta_2\in [0,1]$, $p\in \partial_kF(k_2,c(t^*+h))$ and $q=\frac{\partial F}{\partial c}(k(t^*),c(t^*+\theta_2 h))$ such that
\begin{align*}
&~\rho V'(k_1)(k(t^*+h)-k(t^*))\\
=&~\rho (V(k(t^*+h))-V(k(t^*)))\\
=&~F(k(t^*+h),c(t^*+h))V'(k(t^*+h))+u(c(t^*+h),k(t^*+h))\\
&~-F(k(t^*),c(t^*))V'(k(t^*))-u(c(t^*),k(t^*))\\
=&~(F(k(t^*+h),c(t^*+h))-F(k(t^*),c(t^*+h)))V'(k(t^*+h))\\
&~+(F(k(t^*),c(t^*+h))-F(k(t^*),c(t^*)))V'(k(t^*+h))\\
&~+F(k(t^*),c(t^*))(V'(k(t^*+h))-V'(k(t^*)))\\
&~+u(c(t^*+h),k(t^*+h))-u(c(t^*+h),k(t^*))\\
&~+u(c(t^*+h),k(t^*))-u(c(t^*),k(t^*))\\
=&~pV'(k(t^*+h))(k(t^*+h)-k(t^*))+\dot{k}(t^*)(V'(k(t^*+h))-V'(k(t^*)))\\
&~+\frac{\partial u}{\partial k}(c(t^*+h),k(t^*+\theta_1h))(k(t^*+h)-k(t^*))\\
&~+\left(\frac{\partial u}{\partial c}(c(t^*+\theta_2h),k(t^*+\theta_2h))+qV'(k(t^*+h))\right)(c(t^*+h)-c(t^*))\\
&~+\left(\frac{\partial u}{\partial c}(c(t^*+\theta_2h),k(t^*))-\frac{\partial u}{\partial c}(c(t^*+\theta_2h),k(t^*+\theta_2h))\right)(c(t^*+h)-c(t^*))
\end{align*}
by the mean value theorem.\footnote{See Subsection 3.5.} To modify this equation,
\begin{align}
&~\frac{(\rho V'(k_1)-pV'(k(t^*+h)))(k(t^*+h)-k(t^*))}{h}\nonumber \\
=&~\frac{\dot{k}(t^*)(V'(k(t^*+h))-V'(k(t^*)))}{h}\nonumber \\
&~+\frac{\frac{\partial u}{\partial k}(c(t^*+h),k(t^*+\theta_1h))(k(t^*+h)-k(t^*))}{h}\label{EVA2}\\
&~+\frac{\left(\frac{\partial u}{\partial c}(c(t^*+\theta_2h),k^*(t^*+\theta_2h))+qV'(k(t^*+h))\right)(c(t^*+h)-c(t))}{h}\nonumber \\
&~+\frac{\left(\frac{\partial u}{\partial c}(c(t^*+\theta_2h),k^*(t^*))-\frac{\partial u}{\partial c}(c(t^*+\theta_2h),k^*(t^*+\theta_2h))\right)(c(t^*+h)-c(t^*))}{h}.\nonumber
\end{align}
Because $k(t^*+h)\le k(t^*)<\varepsilon$, $p>\inf_{c\ge 0}D_{k,+}F(k(t^*),c)>\rho$, and thus
\begin{align*}
&~\liminf_{h\downarrow 0}\frac{(\rho V'(k_1)-pV'(k(t^*+h)))(k(t^*+h)-k(t^*))}{h}\\
\ge&~(\rho-\inf_{c\ge 0}D_{k,+}F(k(t^*),c))V'(k(t^*))\dot{k}(t^*)>0.
\end{align*}
Meanwhile, the first and second terms of the right-hand side of (\ref{EVA2}) are always nonpositive. Because of the definition of $c(t)$ and the first-order condition,
\[\frac{\partial u}{\partial c}(c(t^*+\theta_2h),k(t^*+\theta_2h))=-rV'(k(t^*+\theta_2h)),\]
where $r=\frac{\partial F}{\partial c}(k(t^*+\theta_2h),c(t^*+\theta_2h))$, and thus the absolute value of the third term of the right-hand side of (\ref{EVA2}) is bounded from
\begin{align*}
&~\frac{q(V'(k(t^*+h))-V'(k(t^*)))\times |c(t^*+h)-c(t^*)|}{h}\\
&~+\frac{|r-q|V'(k(t^*+\theta_2h))\times |c(t^*+h)-c(t^*)|}{h},
\end{align*}
where,
\[\frac{V'(k(t^*+h))-V'(k(t^*))}{h}\to V''(k(t^*))\dot{k}(t^*)\mbox{ as }h\downarrow 0,\]
\[\frac{r-q}{h}=\theta_2\frac{\partial^2 F}{\partial k\partial c}(k(t^*+\theta'h),c(t^*+\theta_2h))\dot{k}(t^*+\theta'h)\mbox{ for some }\theta'\in [0,\theta_2].\]
Therefore, the third term of the right-hand side converges to zero as $h\downarrow 0$. Finally, because $\frac{\partial u}{\partial c}$ is differentiable in $k$ on $\mathbb{R}^2_{++}$, the fourth term of the right-hand side of (\ref{EVA2}) is
\[\theta_2\frac{\partial^2u}{\partial k\partial c}(c(t^*+\theta_2h),k(t^*+\theta''h))\dot{k}(t^*+\theta''h)(c(t^*+h)-c(t^*))\]
for some $\theta''\in [0,\theta_2]$. Therefore, the fourth term also converges to $0$ as $h\to 0$, and thus the limsup of the right-hand side of (\ref{EVA2}) is nonpositive, which is a contradiction.

Hence, $k(t)\ge \varepsilon$ for every $t\in [0,T]$. By Lemma \ref{Lemma3}, there exists a solution $k^*(t)$ to (\ref{SOL}) defined on $\mathbb{R}_+$. Clearly, $k^*(t)\ge \varepsilon$ for every $t\ge 0$, and thus $\inf_{t\ge 0}k^*(t)>0$. This completes the proof of Step 1.
\end{proof}

\begin{step}
Under the assumptions in Step 1, choose a solution $k^*(t)$ to equation $(\ref{SOL})$ defined on $\mathbb{R}_+$, and define $c^*(t)=c^*(V'(k^*(t)),k^*(t))$. Then, $(k^*(t),c^*(t))\in B_{\bar{k}}$ and
\[V(\bar{k})=\lim_{T\to \infty}\int_0^Te^{-\rho t}u(c^*(t),k^*(t))dt.\]
\end{step}

\begin{proof}[{\bf Proof of Step 2}]
By Step 1, there exists $\varepsilon>0$ such that
\[\varepsilon\le k^*(t)\le k^+(t,\bar{k})\]
for all $t\ge 0$. Because $V$ is a classical solution to the HJB equation,
\[F(k^*(t),c^*(t))V'(k^*(t))+u(c^*(t),k^*(t))=\rho V(k^*(t))\]
for every $t\ge 0$. Therefore,
\begin{align*}
\int_0^Te^{-\rho t}u(c^*(t),k^*(t))dt=&~\int_0^Te^{-\rho t}[F(k^*(t),c^*(t))V'(k^*(t))+u(c^*(t),k^*(t))]dt\\
&~-\int_0^Te^{-\rho t}V'(k^*(t))\dot{k}^*(t)dt\\
=&~-\int_0^T[-\rho e^{-\rho t}V(k^*(t))+e^{-\rho t}V'(k^*(t))\dot{k}^*(t)]dt\\
=&~\int_0^T\frac{d}{dt}[-e^{-\rho t}V(k^*(t))]dt\\
=&~V(\bar{k})-e^{-\rho T}V(k^*(T)),\\
\end{align*}
and thus,
\[V(\bar{k})-e^{-\rho T}V(\varepsilon)\ge \int_0^Te^{-\rho t}u(c^*(t),k^*(t))dt\ge V(\bar{k})-e^{-\rho T}V(k^+(T,\bar{k})).\]
Hence, by (\ref{GC}),
\[\lim_{T\to \infty}\int_0^Te^{-\rho t}u(c^*(t),k^*(t))dt=V(\bar{k}),\]
which implies that $(k^*(t),c^*(t))\in B_{\bar{k}}$. This completes the proof of Step 2.
\end{proof}

\begin{step}
If $(k(t),c(t))\in B_{\bar{k}}$ and $\inf_{t\ge 0}k(t)>0$, then
\[\lim_{T\to \infty}\int_0^Te^{-\rho t}u(c(t),k(t))dt\le V(\bar{k}).\]
\end{step}

\begin{proof}[{\bf Proof of Step 3}]
Suppose that $(k(t),c(t))\in B_{\bar{k}}$ and $\inf_{t\ge 0}k(t)>0$. Because $V$ is a classical solution to the HJB equation,
\begin{align*}
\int_0^Te^{-\rho t}u(c(t),k(t))dt=&~\int_0^Te^{-\rho t}[F(k(t),c(t))V'(k(t))+u(c(t),k(t))]dt\\
&~-\int_0^Te^{-\rho t}V'(k(t))\dot{k}(t)dt\\
\le&~-\int_0^T[-\rho e^{-\rho t}V(k(t))+e^{-\rho t}V'(k(t))\dot{k}(t)]dt\\
=&~\int_0^T\frac{d}{dt}[-e^{-\rho t}V(k(t))]dt\\
=&~V(\bar{k})-e^{-\rho T}V(k(T)).
\end{align*}
Because $\inf_{t\ge 0}k(t)>0$ and $V$ satisfies (\ref{GC}), the right-hand side converges to $V(\bar{k})$ as $T\to \infty$. Therefore,
\[\lim_{T\to \infty}\int_0^Te^{-\rho t}u(c(t),k(t))dt\le V(\bar{k}),\]
as desired. This completes the proof of Step 3.
\end{proof}

Steps 1-3 state that all of our claims in Proposition \ref{Prop4} are correct. This completes the proof.
\end{proof}
\setcounter{step}{0}

The following ODE is used in the proof of the next theorem.
\begin{equation}
\dot{k}(t)=F(k(t),c^*(\bar{V}'(k(t)),k(t))),\ k(0)=\bar{k}.\label{SOL2}
\end{equation}

\begin{thm}\label{Theorem2}
Suppose that Assumptions 1-7 hold. Then, $\bar{V}$ is the unique classical solution to the HJB equation in $\mathscr{V}$. 
\end{thm}

\begin{proof}
Because Assumptions 1-7 hold, $\bar{V}\in \mathscr{V}$, and it is a classical solution to the HJB equation. By Proposition \ref{Prop4}, there exists a solution $k^*(t)$ to (\ref{SOL2}) defined on $\mathbb{R}_+$, where $\inf_{t\ge 0}k^*(t)>0$. Define $c^*(t)=c^*(\bar{V}'(k^*(t)),k^*(t))$. Then, by Proposition \ref{Prop4}, $(k^*(t),c^*(t))\in B_{\bar{k}}$ and
\[\bar{V}(\bar{k})=\lim_{T\to \infty}\int_0^Te^{-\rho t}u(c^*(t),k^*(t))dt.\]
Next, suppose that $V\in\mathscr{V}$ is also a classical solution to the HJB equation. Because $\inf_{t\ge 0}k^*(t)>0$, by Proposition \ref{Prop4},
\[V(\bar{k})\ge \lim_{T\to \infty}\int_0^Te^{-\rho t}u(c^*(t),k^*(t))dt=\bar{V}(\bar{k}).\]
By Proposition \ref{Prop4}, there exists a solution $k^+(t)$ to (\ref{SOL}) defined on $\mathbb{R}_+$ such that $\inf_{t\ge 0}k^+(t)>0$. Define $c^+(t)=c^*(V'(k^+(t)),k^+(t))$. Then, by Proposition \ref{Prop4}, $(k^+(t),c^+(t))\in B_{\bar{k}}$, and thus
\[V(\bar{k})=\lim_{T\to \infty}\int_0^Te^{-\rho t}u(c^+(t),k^+(t))dt\le \bar{V}(\bar{k}).\]
Hence, we conclude that $V=\bar{V}$. This completes the proof.
\end{proof}

Therefore, under Assumptions 1-7, the HJB equation is the perfect characterization for the value function in the functional space $\mathscr{V}$.

In the proof of Theorem 2, if $\int_0^{\infty}e^{-\rho t}u(c^*(t),k^*(t))dt$ is defined in the sense of the Lebesgue integral, then $(k^*(t),c^*(t))\in A_{\bar{k}}$, and thus it is a solution to (\ref{MODEL}). Because $\inf_{t\ge 0}k^*(t)>0$, if $\inf_{t\ge 0}c^*(t)>0$, then $(k^*(t),c^*(t))\in A_{\bar{k}}$. However, this is not so easily verified. The following corollary presents three appropriate sufficient conditions for $(k^*(t),c^*(t))$ to be a solution to (\ref{MODEL}).

\begin{cor}
Suppose that Assumptions 1-7 hold, and $k^*(t)$ is a solution to $(\ref{SOL2})$ defined on $\mathbb{R}_+$ such that $\inf_{t\ge 0}k^*(t)>0$. Define $c^*(t)=c^*(\bar{V}'(k^*(t)),k^*(t))$. Suppose that one of the following three conditions holds.
\begin{enumerate}[1)]
\item $u(c,k)$ is bounded from above or below.

\item $k^*(t)$ is bounded.

\item $\liminf_{k\to \infty}c^*(p,k)>0$ for all $p>0$.
\end{enumerate}
Then, $(k^*(t),c^*(t))$ is a solution to $(\ref{MODEL})$.
\end{cor}

\begin{proof}
By Proposition \ref{Prop4} and Theorem \ref{Theorem2}, $\bar{V}$ is a solution to the HJB equation, $(k^*(t),c^*(t))\in B_{\bar{k}}$, and
\[\lim_{T\to\infty}\int_0^Te^{-\rho t}u(c^*(t),k^*(t))dt=\bar{V}(\bar{k}).\]
Therefore, it suffices to show that $(k^*(t),c^*(t))\in A_{\bar{k}}$.

For 1), if $u(c,k)$ is either bounded from above or below, then $A_{\bar{k}}=B_{\bar{k}}$, and thus this result holds.

For 2), suppose that $k^*(t)$ is bounded. As we proved in Proposition \ref{Prop4}, $\inf_{t\ge 0}k^*(t)>0$. Therefore, the trajectory of $k^*(t)$ is included in some compact set $C\subset \mathbb{R}_{++}$. This implies that $\inf_{t\ge 0}c^*(t)>0$, and thus $e^{-\rho t}u(c^*(t),k^*(t))$ is bounded from below. Hence, $(k^*(t),c^*(t))\in A_{\bar{k}}$ as desired.

For 3), let $0<p_1<p_2$, and $c_i=c^*(p_i,k)$. Then,
\[F(k,c_i)p_i+u(c_i,k)=\max_{c\ge 0}\{F(k,c)p_i+u(c,k)\}.\]
By the first-order condition,
\[\frac{\partial F}{\partial c}(k,c_i)p_i=-\frac{\partial u}{\partial c}(c_i,k).\]
If $c_1\le c_2$, then\footnote{Note that $F$ is decreasing in $c$, and thus $\frac{\partial F}{\partial c}(k,c)<0$.}
\[\frac{\partial u}{\partial c}(c_2,k)=-\frac{\partial F}{\partial c}(k,c_2)p_2>-\frac{\partial F}{\partial c}(k,c_2)p_1\ge -\frac{\partial F}{\partial c}(k,c_1)p_1=\frac{\partial u}{\partial c}(c_1,k),\]
which contradicts the concavity of $u$. Therefore, $c^*(p,k)$ is decreasing in $p$. Define $\varepsilon=\inf_{t\ge 0}k^*(t)$. Then,
\[c^*(t)=c^*(\bar{V}'(k^*(t)),k^*(t))\ge \inf_{k\ge \varepsilon}c^*(\bar{V}'(\varepsilon),k)>0,\]
and thus $u(c^*(t),k^*(t))$ is bounded from below. Hence, again $(k^*(t),c^*(t))\in A_{\bar{k}}$, as desired. This completes the proof.
\end{proof}

Note that, the requirement in Corollary 1 is not strong. For example, suppose that $u(c,k)=au_{\theta}(c)+bu_{\theta}(k)$ for $a,b\ge 0$ and a CRRA function $u_{\theta}$ with $\theta\neq 1$, then $u(c,k)$ is either bounded from above or below, and 1) holds. Next, suppose that $F(k,c)=g(k)-dk-c$, $d>0$, $g$ is continuously differentiable and $\lim_{k\to \infty}g'(k)<d$. Then, $k^+(t,\bar{k})$ is bounded, and thus $k^*(t)$ is also bounded, and 2) holds. Third, if $F(k,c)=f(k)-h(c)$ for a convex function $h$ and $u(c,k)=u(c)$, then $c^*(p,k)$ is independent of $k$. In this case, $\liminf_{k\to \infty}c^*(p,k)>0$ is trivially satisfied, and 3) holds.

It is interesting that the converse of Corollary 1 holds unconditionally.

\begin{cor}
Suppose that Assumptions 1-7 hold and there exists a solution $(k^*(t),c^*(t))$ to $(\ref{MODEL})$ such that $\inf_{t\ge 0}k^*(t)>0$. Then, $k^*(t)$ is a solution to $(\ref{SOL2})$, and $c^*(t)=c^*(\bar{V}'(k^*(t)),k^*(t))$ for almost all $t\ge 0$.
\end{cor}

\begin{proof}
Suppose that $k^*(t)$ violates (\ref{SOL2}). Then, there exists $T>0$ such that the set $\{t\in [0,T]|c^*(t)\neq c^*(\bar{V}'(k^*(t)),k^*(t))\}$ is not a null set. By Theorem 2, $\bar{V}$ is a classical solution to the HJB equation, and thus,
\begin{align}
\int_0^Te^{-\rho t}u(c^*(t),k^*(t))dt=&~\int_0^Te^{-\rho t}[F(k^*(t),c^*(t))\bar{V}'(k^*(t))+u(c^*(t),k^*(t))]dt\nonumber \\
&~-\int_0^Te^{-\rho t}\bar{V}'(k^*(t))\dot{k}^*(t)dt\nonumber \\
<&~-\int_0^T[-\rho e^{-\rho t}\bar{V}(k^*(t))+e^{-\rho t}\bar{V}'(k^*(t))\dot{k}^*(t)]dt\label{EVALL}\\
=&~\int_0^T\frac{d}{dt}[-e^{-\rho t}\bar{V}(k^*(t))]dt\nonumber \\
=&~\bar{V}(\bar{k})-e^{-\rho T}\bar{V}(k^*(T)).\nonumber
\end{align}
By the same arguments, we have that for every $T'\ge T$,
\[\int_T^{T'}e^{-\rho t}u(c^*(t),k^*(t))dt\le e^{-\rho T}\bar{V}(k^*(T))-e^{-\rho T'}(k^*(T')).\]
By (\ref{EVALL}), there exists $\varepsilon>0$ such that
\[\int_0^Te^{-\rho t}u(c^*(t),k^*(t))dt<\bar{V}(\bar{k})-e^{-\rho T}\bar{V}(k^*(T))-2\varepsilon.\]
Choose $\delta>0$ such that $\inf_{t\ge 0}k^*(t)\ge \delta$. Then, for every $T'\ge T$ such that
\[\max\{|e^{-\rho T'}\bar{V}(k^+(T,\bar{k}))|,|e^{-\rho T'}\bar{V}(\delta)|\}<\varepsilon,\]
we have that
\[\int_0^{T'}e^{-\rho t}u(c^*(t),k^*(t))dt<\bar{V}(\bar{k})-e^{-\rho T'}\bar{V}(k^*(T'))-2\varepsilon<\bar{V}(\bar{k})-\varepsilon.\]
Because $T'$ is arbitrary,
\[\int_0^{\infty}e^{-\rho t}u(c^*(t),k^*(t))dt\le \bar{V}(\bar{k})-\varepsilon<\bar{V}(\bar{k}),\]
which is a contradiction. Therefore, $k^*(t)$ is a solution to (\ref{SOL2}). Define $c^+(t)=c^*(\bar{V}'(k^*(t)),k^*(t))$. Then,
\[\dot{k}(t)=F(k^*(t),c^+(t))\]
for all $t\ge 0$, and
\[\dot{k}(t)=F(k^*(t),c^*(t))\]
for almost all $t\ge 0$, which implies that $c^*(t)=c^+(t)$ almost everywhere. This completes the proof.
\end{proof}

We believe that the equation (\ref{SOL2}) is useful for analyzing the model, even if the model is the RCK model with the Inada condition. In the RCK model, the unique solution is characterized by the simultaneous differential equation of the technology constraint (\ref{TC}) and the following Euler equation:\footnote{Note that, the right-hand side of (\ref{SOL2}) is not locally Lipschitz in $k$, because the function $\bar{V}'(k)$ is not necessarily locally Lipschitz. Therefore, in the proof of Corollary 1, the Carath\'eodory-Picard-Lindel\"of uniqueness theorem cannot be applied, and thus we cannot prove the uniqueness of the solution to (\ref{MODEL}). If (\ref{MODEL}) is the RCK model, it is known that the solution is unique. We guess that the solution to (\ref{MODEL}) is unique in a very large class of macroeconomic models, but probably the proof of this conjecture becomes difficult.}
\[\frac{d}{dt}(u'\circ c)(t)=[\rho-f'(k(t))](u'\circ c)(t).\]
In this simultaneous system, the initial value of $c(t)$ is not specified, and this characterization is incomplete. Therefore, to solve the model, we need an additional requirement, and usually, the transversality condition is used as this requirement.

However, there is a problem. In the RCK model, there uniquely exists a steady state, and this is not stable but only \textbf{semistable}. In a semistable system, any simple numerical computation method for the solution such as the Runge-Kutta method does not work well, because the error diverges as $t$ increases. Meanwhile, the equation (\ref{SOL2}) is a simple one-dimensional differential equation, and the steady state is stable. Therefore, the error of numerical calculation converges to zero as $t$ increases. Hence, using (\ref{SOL2}), we can easily obtain an approximate solution using a usual numerical computation method.

\section{Discussions}

\subsection{Example of the Actual Computation}
In this subsection, we provide an example that demonstrates that our results work well.

\vspace{12pt}
\noindent
\textbf{Example} (logarithmic AK model). Let $u(c,k)=\log c$ and $F(k,c)=\gamma k-c$, where $\gamma>\rho$. By Theorem \ref{Theorem2}, $\bar{V}$ is the unique solution to the HJB equation in $\mathscr{V}$. By the first-order condition,
\[c^*(p,k)=\frac{1}{p}.\]
Therefore, the HJB equation is
\[\gamma k\bar{V}'(k)-1-\log \bar{V}'(k)=\rho \bar{V}(k).\]
We guess that $\bar{V}(k)$ is twice continuously differentiable. Thus, to differentiate both sides,
\[\gamma \bar{V}'(k)+\gamma k\bar{V}''(k)-\frac{\bar{V}''(k)}{\bar{V}'(k)}=\rho \bar{V}'(k),\]
and thus, $\bar{V}'(k)$ solves the following differential equation:
\[(\gamma k-(x(k))^{-1})x'(k)=-(\gamma-\rho)x(k).\]
We guess that $(x(k))^{-1}=\rho k$. Then, $x(k)=\rho^{-1}k^{-1}$ and $x'(k)=-\rho^{-1}k^{-2}$. We can easily verify that this function is actually the solution to the above equation. Hence, we obtain
\[\bar{V}'(k)=\rho^{-1}k^{-1},\]
and thus,
\[\bar{V}(k)=\rho^{-1}\log k+C\]
for some constant $C$. Using the HJB equation itself,
\[\gamma \rho^{-1}-1+\log \rho+\log k=\log k+\rho C,\]
and thus,
\[C=\rho^{-1}[\log \rho+\gamma \rho^{-1}-1].\]
In this case, the differential equation (\ref{SOL2}) is
\[\dot{k}(t)=\gamma k(t)-\rho k(t)=(\gamma-\rho)k(t),\ k(0)=\bar{k},\]
and thus, by Corollary 1,
\[k^*(t)=\bar{k}e^{(\gamma-\rho)t},\ c^*(t)=\rho \bar{k}e^{(\gamma-\rho)t}\]
is the solution to the problem (\ref{MODEL}). This completes the calculation.

\subsection{The Magic of Capital}
In this paper, we carefully avoid using $\bar{V}(0)$ in the proofs, although $\bar{V}(0)$ can be defined in the usual manner. Hence, readers may think that the difficulty of the proofs in this paper can be reduced using $\bar{V}(0)$. For example, suppose that $u(0,0)=-\infty$. Then, it is expected that $\bar{V}(0)=-\infty$. In this case, if $(k(t),c(t))\in A_{\bar{k}}$ and $k(T)=0$, then
\[\int_T^{\infty}e^{-\rho t}u(c(t),k(t))dt\le e^{-\rho T}\bar{V}(0)=-\infty,\]
and thus, if $\int_0^{\infty}e^{-\rho t}u(c(t),k(t))dt>-\infty$, then $k(T)>0$ for all $T>0$. This may reduce the difficulty of the proof. Unfortunately, the first expectation is actually incorrect. That is, the conjecture that $\bar{V}(0)=-\infty$ whenever $u(0,0)=-\infty$ is incorrect. Hosoya \cite{HOB} found a counterexample of the above conjecture.

Suppose that $\rho=1,\ F(k,c)=\sqrt{k}-c$ and $u(c,k)=-\frac{1}{\sqrt{c}}$. Then, Assumptions 1-7 hold. Consider a pair $(k(t),c(t))$ defined as follows:
\[k(t)=\frac{t^2}{16},\ c(t)=\frac{t}{8}.\]
We can easily check that $(k(t),c(t))\in A_0$. Moreover,
\begin{align*}
0\ge&~\int_0^{\infty}e^{-\rho t}u(c(t),k(t))dt\\
\ge&~-\sqrt{8}\left[\int_0^1\frac{1}{\sqrt{t}}dt+\int_1^{\infty}e^{-t}dt\right]\\
=&~-\sqrt{32}-\sqrt{8}e^{-1}>-\infty,
\end{align*}
which implies that $\bar{V}(0)>-\infty$.

In the above example, the technology is a variety of RCK technology with $f(k)=\sqrt{k}$. In many macroeconomic textbooks, this function is explained as corresponding to the symmetric Cobb-Douglas technology function in some sense. In this technology, $f(0)=0$, and thus it is thought that if the initial capital stock is absent, then people cannot produce anything. However, the above example shows that actually people can produce something from nothing. We call this phenomenon \textbf{the magic of capital}. Because the above technology corresponds to Cobb-Douglas and the instantaneous utility function is a positive affine transform of a CRRA function with $\theta=\frac{3}{2}$, this magic is a commonplace event.

The magic of capital makes our proofs in this paper difficult. For example, suppose that $(k(t),c(t))\in A_{\bar{k}}$ and $\int_0^{\infty}e^{-\rho t}u(c(t),k(t))dt\ge M$ for some $M$. If $\bar{V}(0)=-\infty$, then we can easily show that $k(t)>0$ for all $t\ge 0$. However, $\bar{V}(0)$ may be greater than $-\infty$, even when $u(0,k)=-\infty$. Therefore, we cannot prohibit $k(t)=0$, even in this case.

Note that, this ``magic'' can be avoided when $F$ is locally Lipschitz around $(0,0)$. However, in applied economics, it is rare that $F$ is locally Lipschitz because, in most cases, it is assumed that $F(k,c)=f(k)-c$ and $f$ satisfies the Inada condition $\mathbb{R}_{++}\subset f'(\mathbb{R}_{++})$. Because $f$ is concave, the Inada condition means that $\lim_{k\to 0}f'(k)=+\infty$, which implies that $F$ is not locally Lipschitz around $(0,0)$. Therefore, we cannot avoid this ``magic'' in economic literature.

\subsection{Comparison with Related Literature}
Usually, the HJB equation is written as a second-order degenerate elliptic differential equation. In fact, in stochastic economic dynamic models, the HJB equation becomes a second-order differential equation. For example, Malliaris and Brock \cite{MB} dealt with this type of equation. Probably, the use of Ito's formula in the middle of the derivation makes the HJB equation second-order. On the other hand, since there is no stochastic variation in the dynamic model considered in this paper, the HJB equation is only a first-order differential equation.

The case in which the HJB equation does not have a classical solution has been highlighted in many studies. Therefore, since Lions \cite{LI}, it has been common to use the viscosity solution as a solution concept in the study of degenerate elliptic differential equations. However, in this paper, we showed that the HJB equation has a classical solution under assumptions that are sufficiently weak in macroeconomic dynamics. It is unclear whether this is also true in other fields, such as models of search theory.

As discussed in the introduction, the HJB equation has mostly been studied in variational problems with a finite time-interval. However, in economics, finite time-interval models are usually undesirable. Many researches that treated the HJB equation with infinite time-interval variational problems assumed that the instantaneous utility function is bounded. This assumption is by far the easiest to use, and dramatically simplifies the proof of our Proposition \ref{Prop3}, for example. However, the most widely used instantaneous utility functions in macroeconomic dynamic models are the class of CRRA functions that include the logarithmic function. Any assumption that excludes this class, regardless of how useful it is, cannot be placed. This is the main reason why the proofs in this paper are by far the most difficult. To the best of our knowledge, the results in this paper are the first theoretical foundations of the HJB equation in macroeconomic dynamics that can be applied to a class of instantaneous utility functions, including the logarithmic function.

Finally, we must mention that the theory of ``weaker'' concept of the solution than viscosity solutions. Consider the following function.
\[H(k,u,p)=\sup_{c\ge 0}\{F(k,c)p+u(c,k)\}-\rho u\]
Then, the HJB equation can be summarized as $H(k,V,V')=0$. In the problems in Section 2, this function $H$ has discontinuity points, which makes the problem difficult. If $H$ can be modified appropriately so that $H$ has good properties, then the value functions in Section 2 may satisfy this ``modified'' HJB equation. For example, Barles \cite{BA} considered such a modification of $H$. Unfortunately, Barles' assumptions are a bit strong that it is not easy to determine whether they can be applied directly to our models. However, this area has been further investigated in recent years, and the theory is developing as applicable to a wider range of problems. See Rampazzo and Sartori \cite{RS} and Motta and Sartori \cite{MS}, and the reference lists therein. However, we do not know whether such a theory on transformed differential equations is effective for application to economics. If further research is developed in economics, it may be shown that the theory of those advanced HJB equations is effective for economic researches.

\section{Concluding Remarks}
In this paper, we presented examples of economic dynamics in which the HJB equation is neither necessary nor sufficient for a function to be the value function. Because this problem is serious, we presented a mathematical foundation of the HJB equation in economic dynamics, and showed that under Assumptions 1-7, the HJB equation provides a perfect characterization of the value function in some class of functions. Moreover, we presented a new method for obtaining the solution to the model using the value function. 

There are several future tasks. First, we want to prove Corollary 1 holds unconditionally. We guess that Corollary 1 holds without any additional assumption, but we cannot prove this yet.

Second, we want to extend our results to some multidimensional models and stochastic models.

Third, we want to obtain a simple method for gaining an approximate solution to the value function. In discrete-time models, there is a famous approximation method that uses Blackwell's inequality and the contraction mapping theorem. We want to obtain a counterpart to this result in a continuous-time model.

Fourth, we need Assumptions 4 and 6 to solve the problem. However, these restrictions are a little strong and any sort of CES function cannot be treated. Hence, we want to relax these constraints.

\section*{Acknowledgements}
The research agenda pursued in this paper was proposed by Hiroyuki Ozaki. We are grateful to him for pointing out an important issue in economic dynamics. This research started from an acute idea proposed by Susumu Kuwata. We received comments from Alexander Zaslavsky and Monika Motta at the 12th AIMS Conference on Dynamical Systems, Differential Equations and Applications. We also received very good advice from Kenji Miyazaki at the 2018 Fall Meeting of the Japanese Economic Association. Keiichi Morimoto also gave us sharp comments at a workshop at Meisei University. Makoto Hanazono made a useful suggestion at the 2020 meeting of the Japanese Society for Mathematical Economics. Additionally, we received many comments and suggestions from Nobusumi Sagara, Toru Maruyama, and Chaowen Yu through private discussions. We would like to express our gratitude to all of them. We also thank the anonymous reviewers for their many suggestions. Of course, the author is responsible for all remaining errors. Finally, this work was supported by JSPS KAKENHI Grant Numbers JP18K12749, JP21K01403.

\end{document}